\documentclass[11pt,a4paper]{emulateapj}

\usepackage{graphicx}
\usepackage{epstopdf}

\usepackage{psfrag}
\usepackage{psfig}
\usepackage{pspicture}
\usepackage{color}

\usepackage{amssymb,amsmath}

\newcommand{\dn}{D$_n$(4000)~}
\newcommand{\hda}{H$\delta_A$~}

\newcommand{\area}{524}

\newcommand{\fuvmag}{\ifmmode{FUV}\else{\it FUV~}}
\newcommand{\nuvmag}{\ifmmode{NUV}\else{\it NUV~}}

\def\apj{ApJ}
\def\apjl{ApJL}
\def\mnras{MNRAS}
\def\pasp{PASP}

\def\aj{AJ}
\def\apjs{ApJS}

\def\nat{Nature}

\slugcomment{Accepted for publication in ApJ}

\shorttitle{Quenching or Bursting: Star Formation Acceleration}
\shortauthors{Martin et al.}

\begin{document}

\title{Quenching or Bursting: Star Formation Acceleration--A New Methodology for Tracing Galaxy Evolution}

\author{
D. Christopher Martin\altaffilmark{1},
Thiago S. Gon\c{c}alves\altaffilmark{2},
Behnam Darvish\altaffilmark{1},
Mark Seibert\altaffilmark{1},
and David Schiminovich\altaffilmark{3},
}

\altaffiltext{1}{California Institute of Technology, MC 405-47, 1200 East
California Boulevard, Pasadena, CA 91125}

\altaffiltext{2}{Observatorio do Valongo, Universidade Federal do Rio de Janeiro,
Ladeira Pedro Antonio, 43, Saude, Rio de Janeiro-RJ 20080-090, Brazil}

\altaffiltext{3}{Department of Astronomy, Columbia University,
New York, New York 10027}

\altaffiltext{4}{Observatories of the Carnegie Institution of Washington,
813 Santa Barbara St., Pasadena, CA 91101}


\begin{abstract}

We introduce a new methodology for the direct extraction of galaxy physical parameters from multi-wavelength photometry and spectroscopy. We use semi-analytic models that describe galaxy evolution in the context of large scale cosmological simulation to provide a catalog of galaxies, star formation histories, and physical parameters. We then apply stellar population synthesis models and a simple extinction model to calculate the observable broad-band fluxes and spectral indices for these galaxies.
We use a linear regression analysis to relate physical parameters to observed colors and spectral indices. The result is a set of coefficients that can be used to translate observed colors and indices into stellar mass, star formation rate, and many other parameters, including the instantaneous time derivative of the star formation rate which we denote the {\it Star Formation Acceleration (SFA)}, We apply the method to a test sample of galaxies with GALEX photometry and SDSS spectroscopy, deriving relationships between stellar mass, specific star formation rate, and star formation acceleration. We find evidence for a mass-dependent SFA in the green valley, with low mass galaxies showing greater quenching and higher mass galaxies greater bursting. We also find evidence for an increase in average quenching in galaxies hosting AGN. A simple scenario in which lower mass galaxies accrete and become satellite galaxies, having their star forming gas tidally and/or ram-pressure stripped, while higher mass galaxies receive this gas and react with new star formation can qualitatively explain our results.

\end{abstract}



\section{Introduction}

There has been growing interest in the nature of the observed
color bimodality in the distribution of galaxies \citep{balogh04,baldry04}, which is
echoed in other galaxy properties \citep{kauffmann03}. The color bimodality is
revealed in a variety of color-magnitude plots, and is particularly dramatic
in the UV-optical color magnitude diagram \citep{wyder07}. The red and blue galaxy concentrations
are commonly denoted the red sequence and the blue ``cloud'', although we
elect to call both concentrations sequences. Recently the blue cloud translated into the specific star formation rate (SSFR)-stellar mass plane is tight enough to be denoted a blue sequence or a ``main-sequence'' for star forming galaxies. Deep galaxy surveys are now
probing the evolution of the red and blue sequences. Work using the COMBO-17 \citep{bell04}, DEEP2 \citep{willmer06,faber07}, and more recently the UltraVISTA \citep{ilbert13} surveys provide evidence that the red sequence has grown in mass
by a factor of three since z$\sim$1. It is natural to ask what processes
have led to this growth, and in particular whether the red sequence
has grown via gas rich mergers, gas-less (dry) mergers, or simple gas
exhaustion. There is also considerable controversy regarding whether AGN feedback has played a role in accelerating
this evolution, with many authors supporting this hypothesis \citep[e.g.,][]{springel05,dimatteo05,maiolino12,olsen13,dubois13,shimizu15}, and many others claiming that energy injection from other feedback mechanisms would dominate the quenching process \citep[e.g.,][]{coil11,aird12}.

Likewise, it has long been assumed that environment plays a significant role in quenching star formation in galaxies. In his seminal work, \citet{dressler80} has shown a strong relation between galaxy morphology and the local density, a relation which translates to an environmental dependency of color and star formation properties on environment \citep[e.g.,][]{zehavi02,balogh04,blanton05,darvish16}. \citet{peng10} have shown that this dependence is stronger for low-mass galaxies -- indicating that quenching of satellite galaxies in clusters is particularly relevant. \citet{peng15} have later argued that the main quenching mechanism in galaxies is ``strangulation'' within clusters. Nevertheless, this results relies on average metallicities and star formation properties of tens of thousands of galaxies, without any regards to processes happening within individual galaxies.

Therefore, we would like very much to identify galaxies which may be in the process of evolving from the blue to the red sequence. \citet{martin07}[M07] have made a first attempt using the \dn and \hda indices as defined in \citet{kauffmann03}, and inferred the total mass flux between both sequences at redshift $z\sim 0.1$. \citet{goncalves12} have extended the analysis to intermediate redshifts ($z\sim 0.8$) and noticed an increased mass flux density at earlier times and and for more massive galaxies, meaning that the phenomenon of star-formation quenching has suffered a sizeable downsizing in the last 6--7 Gyr. Nevertheless, these results rely on the (simplistic) assumption of a star formation history dominated by an exponential decrease in star formation rates in {\it all} green valley galaxies, which cannot be true.

In this paper, we develop a new methodology inspired by earlier work developing simple broad-band and spectral index fitting formulae  \citep{calzetti00,kauffmann03,seibert05,johnson07a,johnson07b} designed to extract physical parameters without explicit SED fitting. Our method starts with model galaxies produced by a semi-analytic model set based on an N-body cosmological simulation (Millennium) ($\S$~\ref{method_models}). 
We then use a linear regression technique to relate photometric and spectral index observables for models binned by the \dn spectral index to model galaxy physical parameters and star formation histories ($\S$ \ref{method_regression}). We define a new star formation history parameter the \textit{Star Formation Acceleration (SFA)} which is the time-derivative of the NUV-i color  ($\S$ \ref{sfa}). A positive SFA corresponds to a galaxy that is quenching (SFR and SSFR dropping) , while a negative SFA indicates a bursting SFR change (SFR and SSFR increasing). We apply this to a matched test ample of SDSS-GALEX galaxies and derive some interesting preliminary results  ($\S$ \ref{applications}).

Throughout this work, we assume a flat $\Lambda$CDM cosmology with $H_{0}$=70 kms$^{-1}$ Mpc$^{-1}$, $\Omega_{m}$=0.3 and $\Omega_{\Lambda}$=0.7. Magnitudes are expressed in the AB system \citep{oke83} and stellar mass and star formation rates are based on a Salpeter initial mass function \citep{salpeter55}.

\section{Method: Galaxy Models} \label{method_models}

One of the principle activities in the field of galaxy evolution is the translation of multi-wavelength photometry and
spectroscopy into galaxy physical parameters.  The vast majority of methods use a spectral energy distribution
(SED) fitting approach. Modelers translate physical parameters and star formation histories into SEDs,
and search for the SED (and corresponding parameters) or range of SEDs which give the best statistical fit. 
Examples of such an approach are given by \cite{kauffmann03} and \cite{salim05,salim07}, who use a Bayesian
analysis of observations fit to a large library of model SEDs that populate galaxy physical parameter space.
The outputs include probability distributions for derived physical parameters.

At the same time, a number of workers have shown that in certain cases simple fitting formulae can 
provide a direct translation of observables into physical parameters. For example,
the UV slope is related to the infrared excess (IRX, the ratio of Far or Total Infrared luminosity to
Far UV luminosity) for starburst \citep{calzetti00}  and normal \citep{seibert05} galaxies.  More complex fitting formulae can be derived
using the \dn spectral index \citep{johnson07a,johnson07b}. \cite{kauffmann03} and these papers
demonstrated that \dn does an excellent job of isolating stellar population age from
other parameters such as extinction. 

This paper introduces a generalization of the fitting formula approach to many physical
parameters and moments of the star formation history. A summary of the approach follows:

\begin{enumerate}

\item We use a semi-analytic model \citep{delucia06} linked to the Millennium cosmological N-body simulation \citep{springel05b}
to provide a large sample of galaxies, star formation histories, and associated physical parameters.

\item We use a simple extinction model and stellar population synthesis code to translate the star formation histories
into observable broad-band fluxes and spectral indices.

\item We bin model SEDs by \dn to remove the principle source of variation, stellar population age.

\item Within each \dn bin we perform a linear regression fit between model physical parameters and
the multiple observables (colors and spectral indices) for the complete galaxy sample. 
We find in general linear (in the log) relationships between the two over a large dynamic range. Fit dispersion
varies with physical parameter and with the collection of available observables. 

\item The matrix of regression coefficients can be used to translate observables into physical parameters (after introducing some offsets),
and to derive observable influence functions, degeneracies and error propagation matrices.

\end{enumerate}

\subsection{Cosmological Simulation and Semi-analytic Model}

We use a set of 24,000 model galaxies produced by the \cite{delucia06} semi-analytic model (SAM) applied to the
Millennium cosmological simulation. Galaxies are modeled in 63 time steps of $\sim$300 Myr each over the redshift
range ($0<z<6$). 

The Millennium simulation \citep{springel05b} is a N-body simulation that follows 2160$^3$ particles since redshift $z=127$ in a cosmological volume 500 $h^{-1}$ Mpc on a side. Assuming a cold dark matter cosmology, it provides a framework in which one can follow the formation of dark matter haloes and the large-scale structure on cosmologically significant scales. \citet{delucia06} used this framework and applied a semi-analytic model which, following dark matter haloes even after accretion onto larger systems, assumed a star formation law that depended on the cold gas mass and a minimum critical value of gas surface density above which new stars were allowed to form. With the addition of active galactic nuclei (AGN) feedback, the authors are able to reproduce the observed trend of short formation time-scales of the most massive elliptical galaxies \citep[e.g.,][]{thomas05}. 

We used 24,000 galaxies (at snapnum=63 or z=0) from
the volume range ($0<x<65$ Mpc, $0<y<65$ Mpc, $0<z<65$ Mpc), where $x$, $y$, and $z$ are the galaxy coordinates in the
Millennium catalog, and absolute magnitude $M_r<-17$. 
Each z=0 galaxy is the base of a merger tree. Each tree and all galaxy predecessors was loaded,
giving a total of 900,000 galaxy models over all 63 time steps and over the redshift range ($0<z<6$). 
All regression fits given below use {\it all} galaxies in {\it all} time-steps (subdivided
only by \dn and in 9 course redshift bins), using rest-frame observables. {\it Hence, all results given below can be applied
to galaxies at any redshift, using k-corrected observables.} 

\subsection{Spectral Energy Distributions}

\subsubsection{Stellar Population Synthesis}\label{sec_pop_synthesis}

We use the SAM model star formation rate for each galaxy
and the merger tree to calculate a star formation history for each galaxy at each time step/redshift.
The star formation rate (SFR) vs. time is calculated at each time step and is the sum of the
star formation histories of all predecessor galaxies in the merger tree.
Updated single-stellar population (SSP) stellar population synthesis models of \citet[][CB07]{bruzual03} are used to predict
broad-band luminosities and spectral indices. These models are available in
seven metallicity bins. In each time step, a SSP is created associated with the SFR
and time interval in that time step. The metallicity of the SSP in this time step is
derived from the gas phase metallicity from the SAM (using the closest available
SSP model). We use a Salpeter initial mass function. 

\subsubsection{Dust Extinction}

We have used a simple geometric model for dust extinction.
The SAM predicts gas phase metallicity ($Z_{gas}$), gas mass ($M_{gas}$), and
galaxy size ($r_{gal}$).  We assume that gas and dust are distributed in a uniform
absorbing slab with selective extinction $E_{B-V}$ given by
\begin{equation}
E_{B-V} = C_0 \mu Z_{gas} M_{gas} {r_{gal}}^{-2} ({\cos{i}})^{-1}
\end{equation}
where $C_0$ is a constant (obtained by using the Milky Way values) and $i$ is the galaxy inclination. The constant $\mu$
allows for a larger absorption for young stars than for evolved stars \citep{calzetti94}.
We use $\mu=1$ for stars younger than 10 Myr and $\mu=0.5$ for older stars. 

We use three possible extinction-law models. 1) The starburst extinction law from \cite{calzetti00} gives the usual A$_{FUV}$ or IR-excess (IRX) vs. UV slope with A$_{FUV}$ and IRX increasing with $\beta$, the slope of the SED in the FUV/NUV region. 
2) The Milky Way extinction law from \cite{cardelli89} has an IRX-$\beta$ relationship that is flattened and even reversed because of the 2200\AA~bump. 3) A mixed extinction model in which a fraction $f_M$ of the dust follows Milky Way extinction, and a fraction
$1-f_M$ follows the starburst extinction, where $f_M$ is chosen randomly over a range $0<f_M<f_{M,max}$. 
For the results given below we use this third method, which gives a fitting error of 0.3 magnitude for $f_{M,max}=0.5$ rising to 0.5 magnitude for $f_{M,max}=1$. In order to incorporate the positive definite quantity A$_{FUV}$ as a derived parameter, we fit the quantity $IRX_{FUV} = log_{10} (10^{0.4A_{FUV}}-1)$.

\subsubsection{Nebular Emission}

We do not incorporate nebular emission in this version of the model.
In a future paper we will incorporate emission lines and examine 
additional physical parameters that these trace, including SFR and IMF.

\subsubsection{Sample Star Formation Histories}

\begin{figure}
\plotone{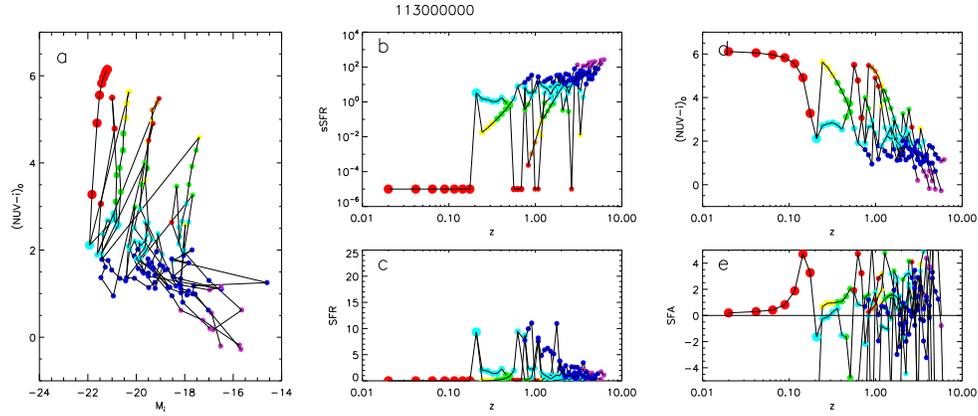}
\caption{Simulated star formation history for a galaxy most like a massive quiescent at the present day. 
a. Color-magnitude diagram (extinction corrected NUV-i vs. M$_i$) showing evolutionary tracks of all the galaxies that eventually merge into the single quiescent galaxy. The final merger occurs at z=0.2. 
Circle size is keyed to $log(M_*)$, and color is keyed to specific star formation rate. b. Specific SFR (sSFR) vs. redshift for all constituent and final galaxy. c. SFR vs. redshift plotted as in panel b. d. NUV-i vs. redshift. e. Star Formation Acceration (cf. \S\ref{sfa}) vs. redshift.
\label{fig_gal_ellipt}}
\end{figure}

\begin{figure}
\plotone{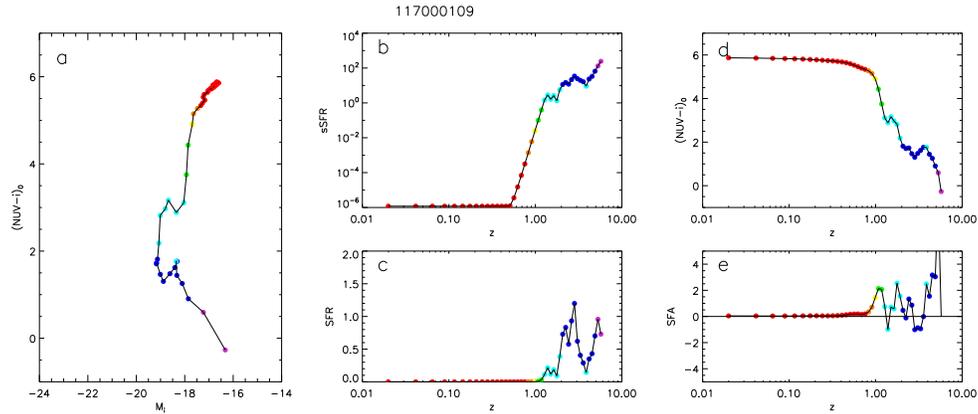}
\caption{Simulated star formation history for a galaxy starting a slow quench at $z\sim 2$. 
a. Color-magnitude diagram (extinction corrected NUV-i vs. M$_i$).
Circle size is keyed to $log(M_*)$, and color is keyed to specific star formation rate. b. Specific SFR (sSFR) vs. redshift for all constituent and final galaxy. c. SFR vs. redshift plotted as in panel b. d. NUV-i vs. redshift. e. Star Formation Acceration (cf. \S\ref{sfa}) vs. redshift.
\label{fig_gal_slowquench}}
\end{figure}

\begin{figure}
\plotone{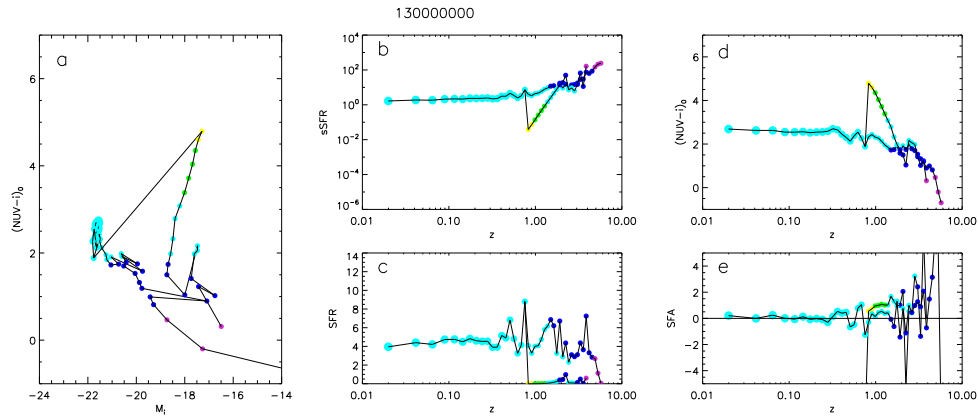}
\caption{Simulated star formation history for a galaxy like a star-forming with a late minor merger at $z\sim0.8$. 
a. Color-magnitude diagram (extinction corrected NUV-i vs. M$_i$) showing evolutionary tracks of all the galaxies that eventually merge into the single galaxy. 
Circle size is keyed to $log(M_*)$, and color is keyed to specific star formation rate. b. Specific SFR (sSFR) vs. redshift for all constituent and final galaxy. c. SFR vs. redshift plotted as in panel b. d. NUV-i vs. redshift. e. Star Formation Acceration (cf. \S\ref{sfa}) vs. redshift.
\label{fig_gal_spiral}}
\end{figure}

\begin{figure}
\plotone{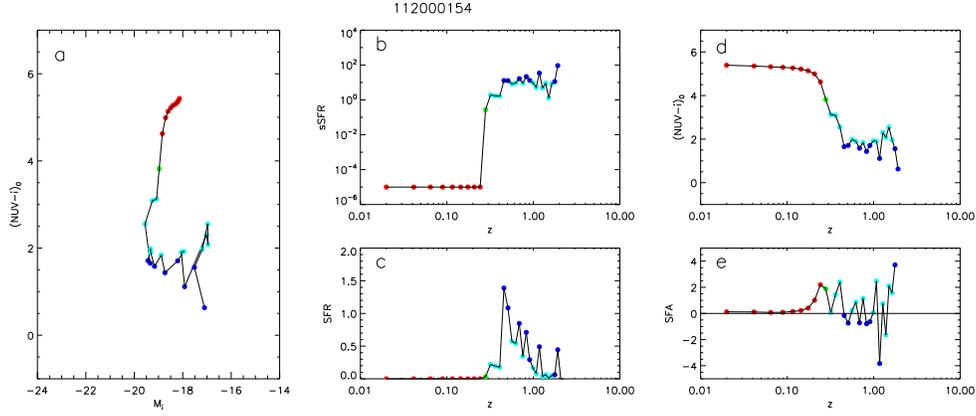}
\caption{Simulated star formation history for a galaxy starting fast quench at $z\sim 0.5$. 
a. Color-magnitude diagram (extinction corrected NUV-i vs. M$_i$) showing evolutionary tracks the galaxy.  Circle size is keyed to $log(M_*)$, and color is keyed to specific star formation rate. b. Specific SFR (sSFR) vs. redshift for all constituent and final galaxy. c. SFR vs. redshift plotted as in panel b. d. NUV-i vs. redshift. e. Star Formation Acceration (cf. \S\ref{sfa}) vs. redshift.
\label{fig_gal_fastquench}}
\end{figure}

\begin{figure}
\plotone{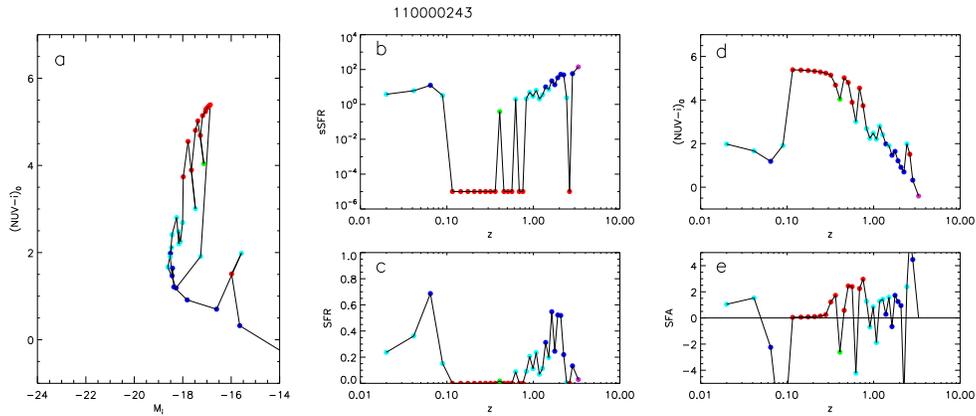}
\caption{Simuated star formation history for a dwarf galaxy quenching at early times and then bursting at $z\sim0.07$ 
a. Color-magnitude diagram (NUV-i vs. M$_i$ including extinction) showing evolutionary tracks of galaxy. Circle size is keyed to $log(M_*)$, and color is keyed to specific star formation rate. b. Specific SFR (sSFR) vs. redshift for all constituent and final galaxy. c. SFR vs. redshift plotted as in panel b. d. NUV-i vs. redshift. e. Star Formation Acceration (cf. \S\ref{sfa}) vs. redshift.
\label{fig_gal_dwarf}}
\end{figure}

In Figures \ref{fig_gal_ellipt}-\ref{fig_gal_dwarf}, we show sample star formation histories and evolution in the NUV-i vs. $M_i$ color-magnitude diagram
from five galaxies (a massive quiescent, a galaxy slow-quenching at $z\sim 2$,
a disk galaxy with relatively constant SFR, a galaxy fast-quenching at $z\sim 0.5$, and a low mass recent starburst galaxy).

\section{Method: Galaxy Physical Parameters} \label{method_regression}

\subsection{Mathematical Motivation}

We would like to recover measures of recent star formation history (SFH) that are non-parametric. Our technique relies on linearization, effectively Taylor expansion to the linear term of a multi-dimensional non-linear function around fixed points. There is a complex, non-linear relationship between observed colors and spectral indices and physical parameters. For a Single Stellar Population (SSP) the principal source of variation is age. A robust measure of SSP age is the spectral index \dn, since extinction has almost no effect (metallicity has some effect, and we defer discussion of this until \S\ref{sec_discussion}). We make an ansatz that once \dn is specified, there is a linear relationship between observable colors and indices and star formation metrics such as SFR, specific SFR, stellar mass, and recent changes in SFR. This relationship can be tested with a family of star formation histories and a stellar population synthesis models, as long as this family spans the space of real galaxy star formation histories.  We also assume that physical parameters such as stellar and gas metallicity, gas mass, and extinction also have this linear relationship with observables. Testing this requires relating the star formation histories to the physical parameters with for example a semi-analytic model connected to a realistic cosmological simulation. 

\subsection{Regression Method and Star Formation Acceleration Parameter} \label{sfa}

We use standard multiple linear regression (MLR) to relate physical parameters to observed
properties. For this initial work, we use the following observables. All samples are binned in \dn
with $\Delta$\dn=0.05. Other observables used in
this initial study are the colors: FUV-NUV, NUV-u, u-g, g-r, r-i; the spectral index \hda, and the absolute magnitude $M_i$. FUV and NUV are GALEX
bands, and u,g,r,i,z are SDSS bands. 

We perform MLR between all of these observables and each of the following physical parameters:
stellar mass ($\log{M_*}$), star formation rate ($\log{SFR}$), FUV extinction ($A_{FUV}$), 
extinction correction to NUV-i ($\Delta(NUV-i)=(NUV-i)_0-(NUV-i)_{obs}$ where $(NUV-i)_0$ is the extinction-corrected $NUV-i$, mass-weighted
stellar age ($\log{t_*}$), gas mass ($\log{M_{gas}}$), gas metallicity ($Z_{gas}$), and stellar
metallicity ($Z_*$).

We also fit two additional functions related to moments
of the star formation history. We call the ``Star Formation Acceleration (SFA)'' the
time derivative of the extinction-corrected NUV-i color ($SFA\equiv d(NUV-i)_0/dt$) (note that the SFA defined using NUV-r vs. SFA defined using NUV-i differ by only 1\%). The SFA is calculated
using the current and previous time steps, and is quantified as mag Gyr$^{-1}$. In the lowest redshift bin ($0<z<0.3$), applicable in the results we present below, the time steps are separated by 0.3 Gyr. 

While there are several possible definitions one could use for SFA (specifically, $d(SFR)/dt$, $d(sSFR)/dt$, $d\log(sSFR)/dt$, and $d(NUV-i)_0/dt$), we have chosen to use the latter for the following reasons. 1) $d(SFR)/dt$ is not mass normalized and will scale with galaxy mass, making direct comparisons between mass bins less informative. 2) $sSFR$ can vary over many orders of magnitude making comparisons of galaxies in different $sSFR$ bins less informative. 3) $d\log(sSFR)/dt$ is more useful and can track changes across the CMD. But it can take on large negative and even indefinite values when quenching occurs rapidly that can only be bounded by using arbitrary parameters to limit the change.  We have experimented with using log(sSFR), finding that the fits are slgnificantly worse than with our adopted definition ($2.5\sigma(SFA_{sSFR})=4.4$ vs. $\sigma(SFA_{NUV-i})=1.5$). 4) Our adopted definition is logarithmic and $(NUV-i)_0$ is well correlated with log sSFR (with $(NUV-i)_0 \simeq Const. -2.5 \log sSFR$ for $1<(NUV-i)_0<5$ with a break and a slight shallower function for $(NUV-i)_0>5$). It is also well behaved even with abrupt changes in SFR and sSFR. Because it and all the observed colors and spectral indices are light-weighted moments of the star formation history they are better correlated and the fit dispersions much lower. Finally, using this definition we can make a direct comparison to our previous work calculating the mass flux across the color-magnitude diagram. This approach will ultimately be used to tie together different epochs of the observed CMD by comparing the measured CMD flux to the measured CMD changes with redshift.

We also calculate a past SFA as well (the SFA for the two time-steps preceding the current one
(SFJ).  Note that the SFA and SFJ can be positive or negative. A negative SFA would signal
a recent starburst, while a positive SFA would indicate on-going star-formation quenching. 
A positive SFA and SFJ would indicate a longer-duration quench. 

\begin{figure}
\plotone{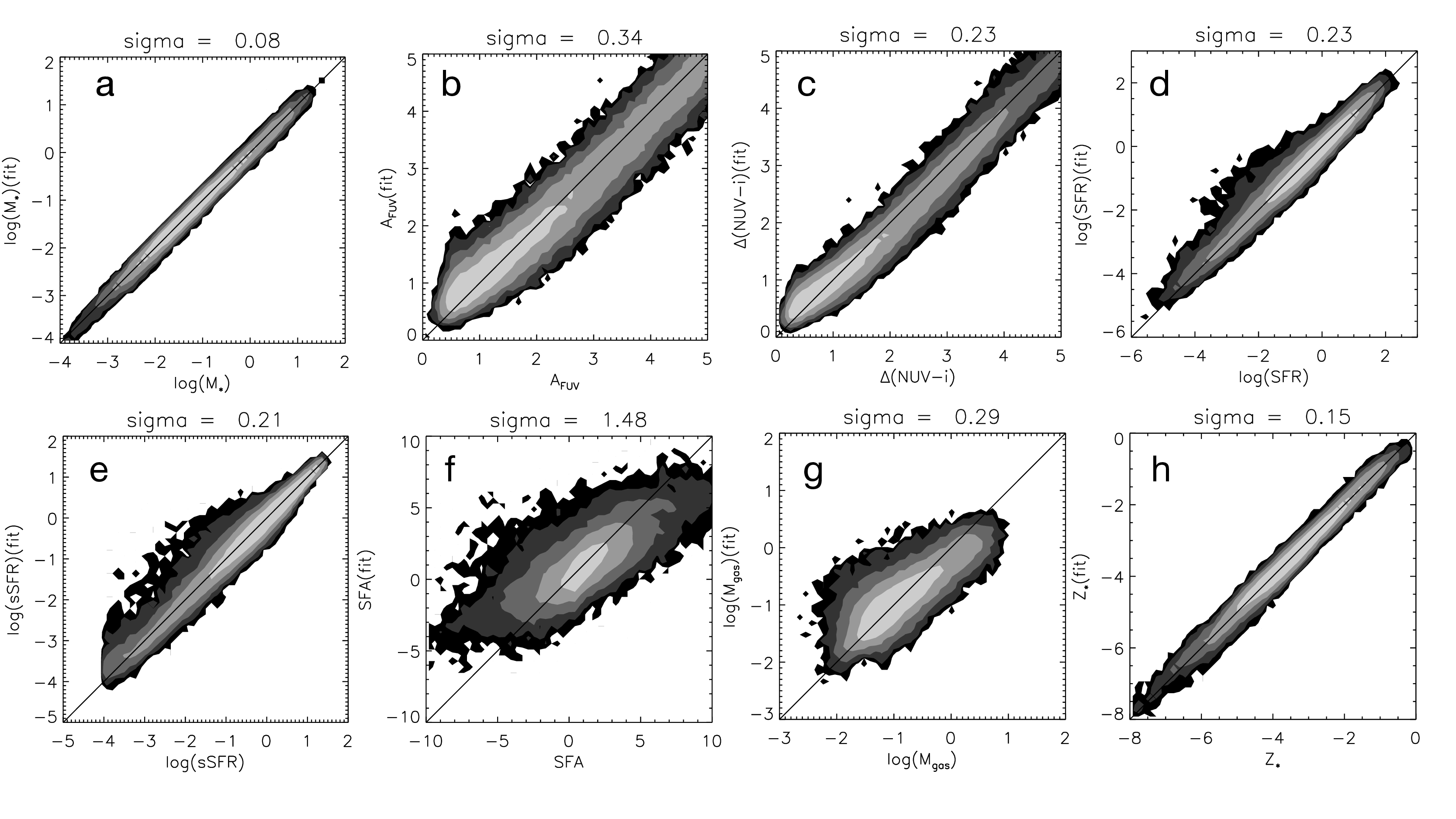}
\caption{Results of using linear regression parameter recovery compared to actual model parameter. For each parameter the model parameter is plotted on the abscissa and recovered parameter using observables (fit) plotted on the ordinate. We include all valued of D$_n$(4000), all galaxies, and all redshifts. The rms deviation of the fit parameter from the input parameter is given above each panel. a. Stellar mass. b. FUV extinction $A_{FUV}$. c. Unextincted NUV-i. 
d. Star formation rate (SFR). e. Specific star formation rate (sSFR). f. Star formation acceleration (SFA) or d(NUV-i)/dt in magnitudes per Gyr. g. Gas mass ($M_{gas}$). h. Stellar  ($Z_*$). 
\label{fig_samplefits}}
\end{figure}

The result is a matrix relating 8 observables to the 10 physical parameters for each of 20 bins in \dn, and in 9 redshift bins
or 20 $8\times11$ matrices. The matrix elements are denoted $M_{p,o,d,z}$ where $p$ refers to the physical
parameter, $o$ to the observable,  $d$ to the \dn value, and $z$ the course redshift bin. In Figure \ref{fig_samplefits} we show
some sample fits combined for all \dn and redshift bins. We note that there is moderate error in the SFA fit as well as some bias. Fitting error is included in assessing the error in our mean SFA calculations. Biases are small and discussed in Appendix \S\ref{sec_model_comp}.


Physical parameters are derived from
\begin{equation}
P_p(est)=\sum_{o=1}^{o=8} M_{o,p,d,z} O_o 
\end{equation}
or for the observable set used here,
\begin{eqnarray}
\notag
P_p(est) & =  & M_{1,p,d,z} (FUV-NUV) +  
\\  
\notag
& & M_{2,p,d,z} (NUV-u) + M_{3,p,d,z} (u-g) +  
\\
\notag
& &  M_{4,p,d,z} (g-r) + M_{5,p,d,z} (r-i) +
\\
 \notag
& & M_{6,p,d,z} D_n(4000) + M_{7,p,d,z} H_{\delta A} +
\\
& & M_{8,p,d,z} M_i + constant
\end{eqnarray}

\subsection{Influence Functions}

In general not all observables used in the above fits are available. Some, such as \hda, may be difficult to obtain.
It is useful therefore to quantify the impact each observable has on each derived physical parameter.
We do this by calculating the relative decrease in variance when using the observable to that when not
using the observable. This is normalized to the total variance in the physical parameter
over the full sample in a given \dn bin:
\begin{equation}
I[p,o,d,z]\equiv {{[<{\sigma_{[p,\bar{o},d,z]}}^2>-<{\sigma_{[p,o,d,z]}}^2>]} \over {{\sigma_{[p,d,z]}}^2}}
\end{equation}
where for physical parameter $p$, \dn bin $d$, and observable $o$ either used $o$ or not used $\bar{o}$.
The mean is taken over all possible non-trivial combinations of observables (with or without observable $o$).
A value of 1.0 would mean that the observable completely eliminates the parameter variance 
when introduced, and a value 0.0 means the observable has no influence on the fit.

For example, for \dn=1.40, the influence function for $A_{FUV}$ is (0.24, 0.34, 0.30, 0.30, 0.49, 0.07, 0.10)
for (FUV-NUV, NUV-u, u-g, g-r, r-i, \hda, $M_i$). Each photometric color makes a contribution to the
fit variance reduction, with NUV-i  reducing over 50\% of the variance. 
Specific SFR ($\log{sSFR}$, or the $\log{SFR/M_*}$, has influence functions
(0.54,0.29,0.06,0.08,0.20,0.17,0.20). The bulk of the information comes from FUV-NUV and NUV-u, with
virtually no impact from u-g or g-r. Finally, SFA has influence functions (0.19,0.16,0.03,0.11,0.11,0.49,0.08).
Most of the information comes from \hda. 

\begin{figure}
\plotone{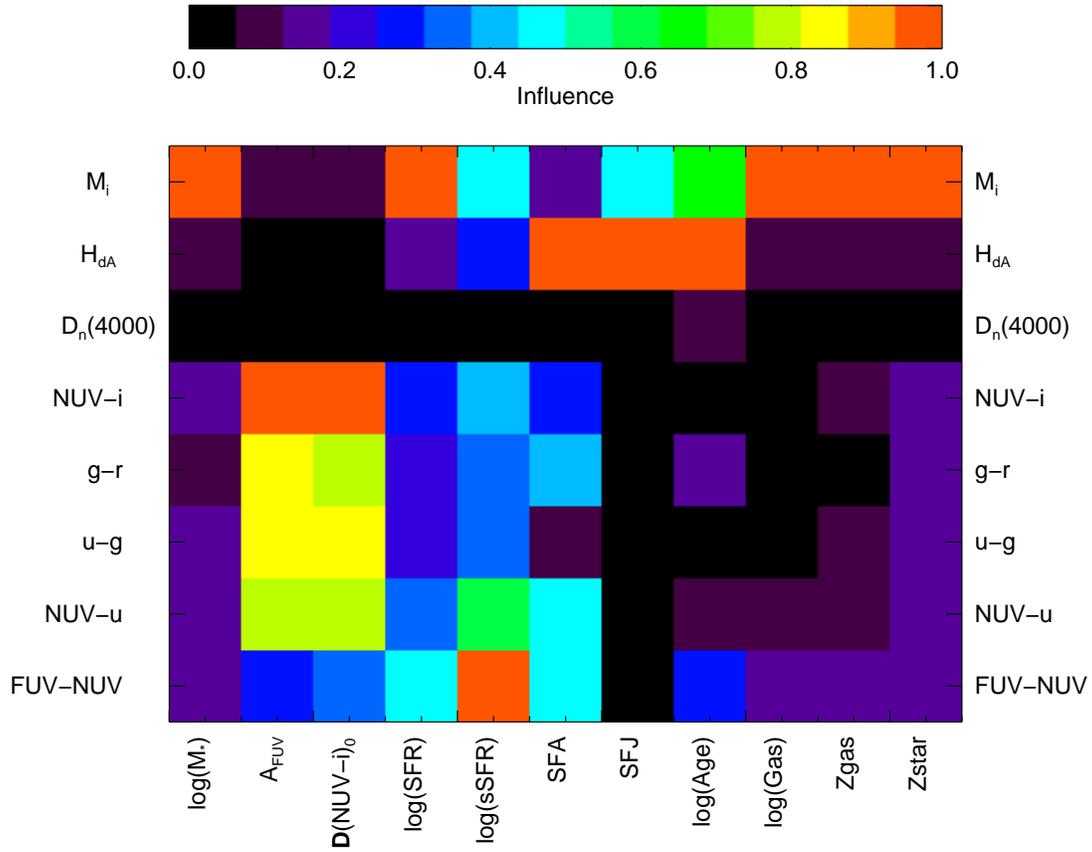}
\caption{Influence functions for each parameter. This gives a graphic representative of the sensitivity of a given observable on recovering a given physical parameter (see text). The influence functions are normalized for each observable so that the observable with the maximum influence is 1.0. For example, $M_i$ has a strong influence on log ($M_*$), log SFR, log $M_{gas}$, Z$_{gas}$, and Z$_{*}$. Extinction and extinction corrected NUV-i  [D(NUV-i) = (NUV-i)$_0$-(NUV-i)] are strongly influenced by NUV-i, g-r, u-g, NUV-u, and FUV-NUV.
\label{fig_influence}}
\end{figure}

Table \ref{tab_influence} gives the mean influence functions (averaged over all \dn).
Figure \ref{fig_influence} shows a color-coded display of the same information.

\subsection{Degeneracies/Observational Basis}

\begin{figure}
\plotone{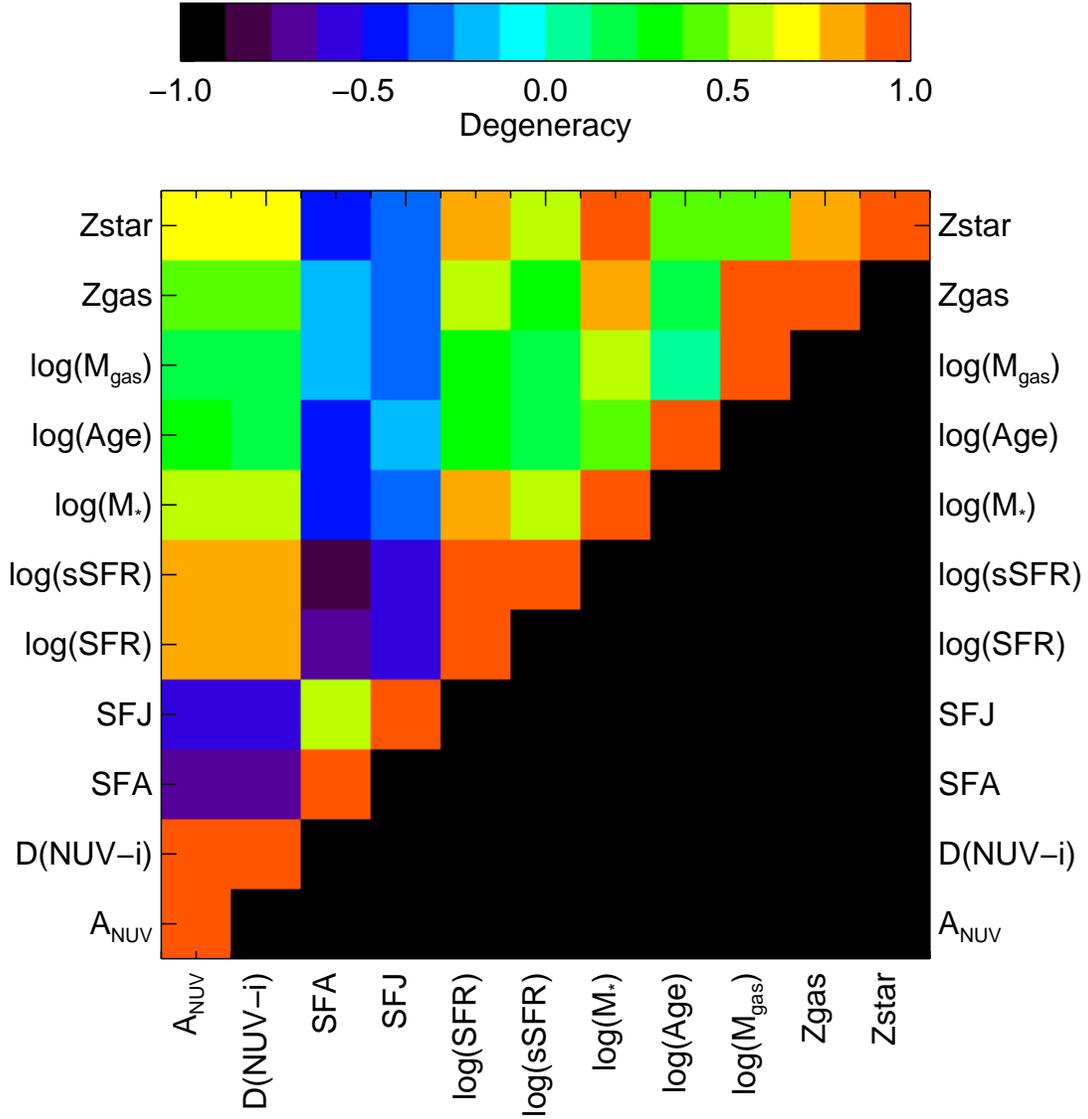}
\caption{Degeneracy between derived physical parameters. A negative degeneracy implies the parameters are inversely correlated. This gives a graphic representative of the degeneracy $D_{i,j}$ between parameter $i$ and $j$ (see text). A degeneracy of $D_{i,j}=1$ or $D_{i,j}=-1$ implies that the parameters cannot be independently extracted. For example, log ($M_*$) and Z$_{*}$ are highly degenerate, as is the change extinction correction to NUV-i [D(NUV-i) = (NUV-i)$_0$-(NUV-i)] and $A_{FUV}$.
\label{fig_gal_degeneracy}}
\end{figure}

This method allows us to quantify parameter degeneracies in a simple fashion.  Consider the 7-dimensional space of observations, and
a single physical parameter $P_i$. A vector exists in this space in the direction that produces the
maximum change in derived physical parameter. This is just the gradient in $P_i$ which is given
by the matrix coefficients:
\begin{equation}
\nabla \vec{P_i} = \sum_o M_{i,o} \hat{j_o}.
\end{equation}
where $\hat{j_o}$ is a unit vector in the direction of the observable $o$ in this multi-dimensional space. The degeneracy of two physical parameters $P_i$ and $P_j$ can be determined from the dot-product of
these two gradients:
\begin{equation}
D_{i,j} \equiv { {\nabla \vec{P_i}  \bullet \nabla \vec{P_j}} \over {|\nabla \vec{P_i}| |\nabla \vec{P_j}|}}
\end{equation}
A degeneracy of $D_{i,j}=1$ would mean that the two derived physical parameters come from the same
linear combination of observables and are completely degenerate. Degeneracy can be negative, if two observables give the same information but with opposite dependencies. The degeneracies averaged over \dn~and redshift bins are shown in Figure \ref{fig_gal_degeneracy}.

We note for example that the mass-weighted age have a degeneracy of -0.72, since both depend strongly on \hda (and \dn). This means that their influence vectors are $\sim$45$\deg$ apart. So while they are related they are not identical. It is not surprising there is degeneracy here. It may be counter-intuitive that the degeneracy is negative, since one associates bursting with younger populations. However, this makes sense. The coefficients are calculated in a fixed \dn bin. A galaxy with a smooth SFR will have a particular \hda associated with that \dn. If there was more SFR in the past (quenching), \hda will be higher than this smooth baseline since it peaks at hundreds of Myr. The mass-weighted age will also be younger. If there was less SFR in the past (e.g., more in the present, bursting) then \hda will be lower than the baseline and the mass-weighted age will be older. 

In some sense all colors and spectral indices are ``light-weighted ages'' with different averaging kernals. For example, extinction-corrected NUV-i is highly correlated with sSFR, since NUV tracks SFR (short-term light-weighted age) and i-band has a very long averaging kernal and therefor is a stellar mass tracer. Thus SFA is derived from color/index differences (see plots in \S\ref{sec_color_correct}) that can be linearized within individual \dn bins.

\subsection{Error Propagation/Observable Figure of Merit}

Since the derived parameters are linear functions of the observables, it is a simple matter to propagate observational errors to determine the total observational error component of the derived parameters. This can then be combined with the fitting error derived from the MLR step. If the observational error is large, and its influence is small, including the observation will actually increase the uncertainty of the derived parameter. Clearly the criterion for including an observable $o$ with an observational error $\sigma_o$ is:
\begin{equation}
{\sigma_{[p,{o},d,z]}}^2 + {M_{o,p,d,z}}^2 {\sigma_o}^2 < {\sigma_{[p,\bar{o},d,z]}}^2
\end{equation}

\section{Applications: GALEX/SDSS Galaxies}  \label{applications}

Once we determine the matrix of linear coefficients, we can proceed to apply the method to real galaxies. We present this simply as an illustration of the potential of the methodology presented in this work, and expect that the full scientific yield will be realized over a range of studies and applications in the future.

\subsection{Observed Sample}

We use the same GALEX/SDSS-spectroscopic sample as in \cite{martin07}.
Our sample is NUV selected in the GALEX Medium
Imaging Survey \citep[MIS;][]{martin05}. The MIS/SDSS DR4 co-sample occupies
\area~ sq. deg. of the north galactic polar cap and the southern
equatorial strip.  Our sample is cut as follows:
1) NUV detection, nuv\_weight $>$ 800;
2) SDSS main galaxy sample, $z_{conf}>0.67$ and specclass=2;
3) $14.5<r_0<17.6$, $16<NUV<23.0$;
4) nuv\_artifact$<$2;
5) field radius less than 0.55 degrees;
6) $0.02<z<0.22$.
We use \dn and \hda as calculated and employed for the SDSS
spectroscopic sample by \cite{kauffmann03} and available as the MPIA/JHU DR4 Value-Added
Catalog. \hda is corrected for nebular emission. 
The sample properties, galactic extinction and k-correction, and cuts are discussed further in M07. 

There are slight differences in the mean colors of the observed sample with respect to the model colors. These are typically $\sim 0.1$ magnitude but rise to $\sim 0.6$ magnitudes in the case of NUV-u for several bins in \dn. Also, model \hda are higher than observed \hda by about 0.5 over a range of \dn.  Model color dispersions are comparable to the observed dispersions when observation errors are included.  The model mean colors in each \dn bin have been adjusted to match the observed mean colors prior to model fitting in order to ensure that the range of derived parameters is not outside the bounds of the fitted parameters. Please see Appendix \ref{sec_color_correct} for further details. 

\begin{figure}
\plottwo{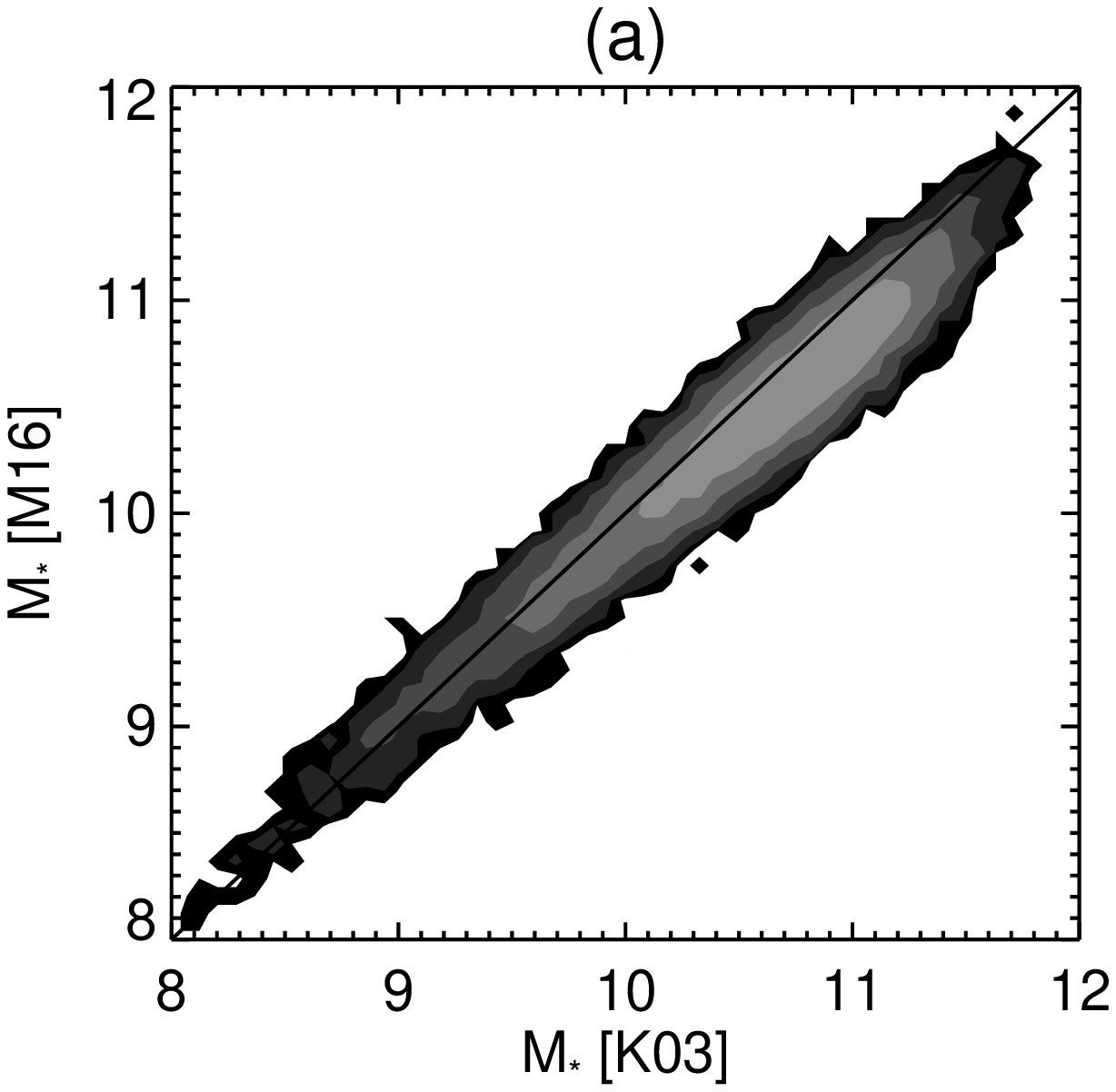}{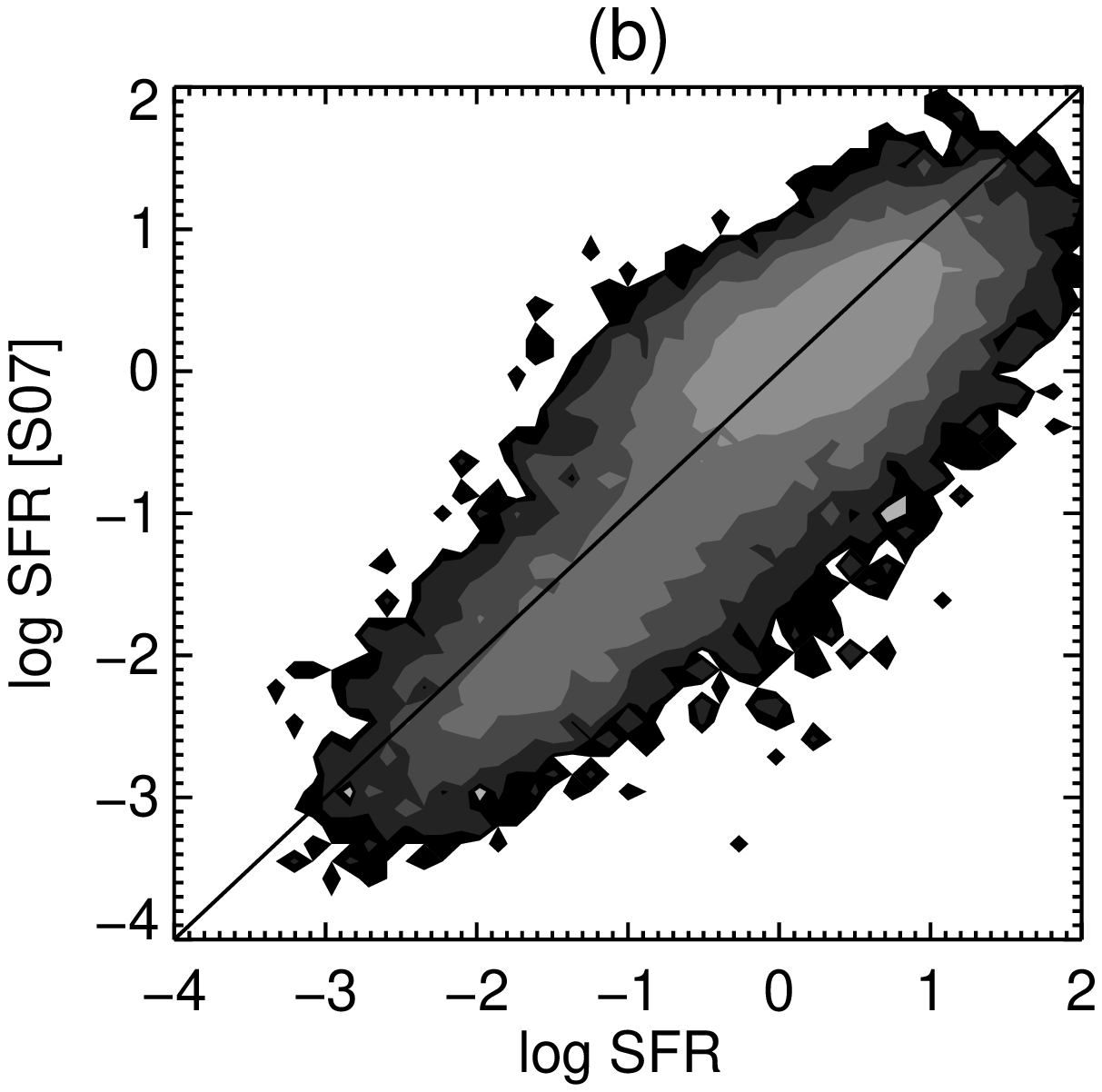}
\plottwo{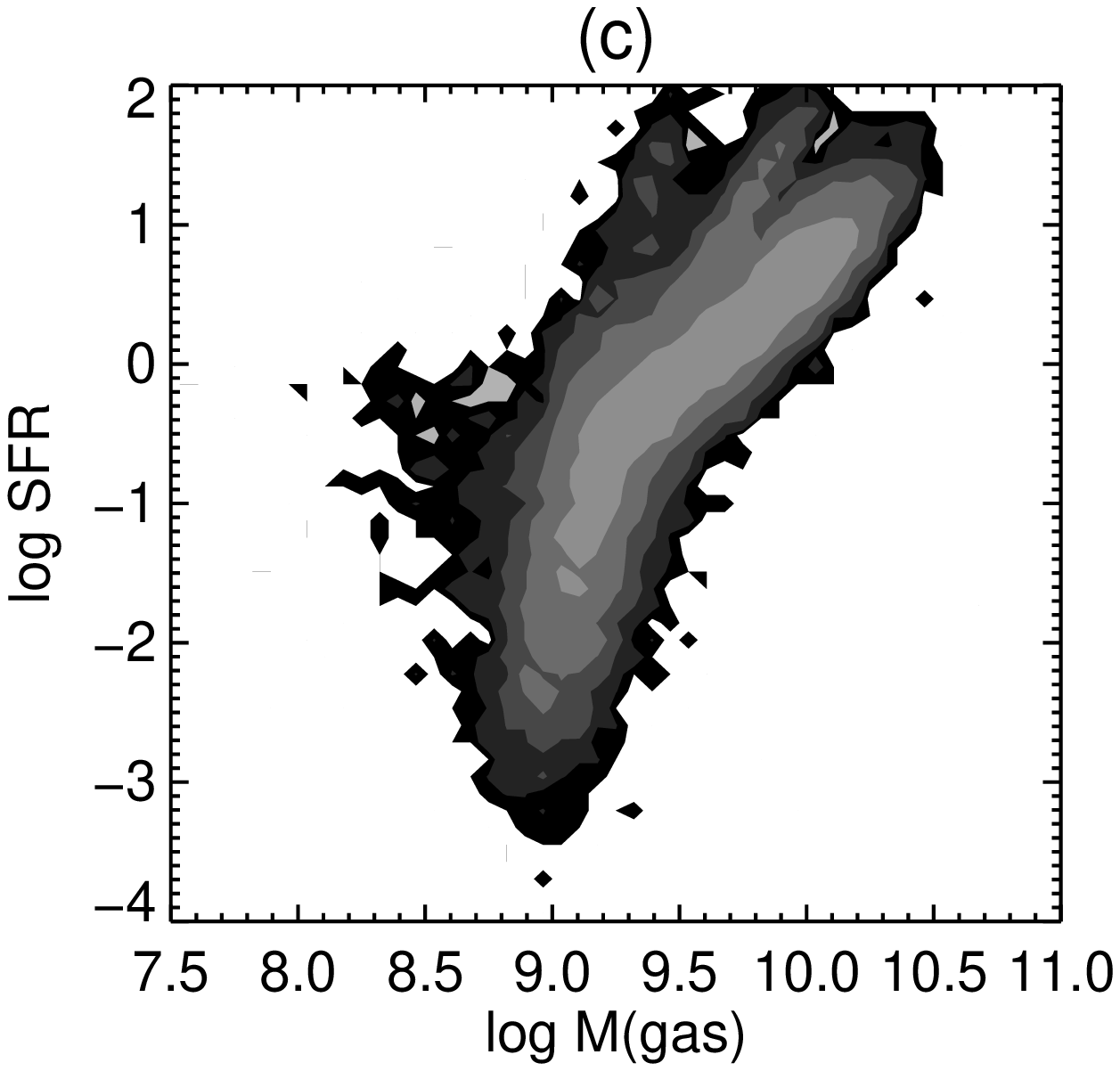}{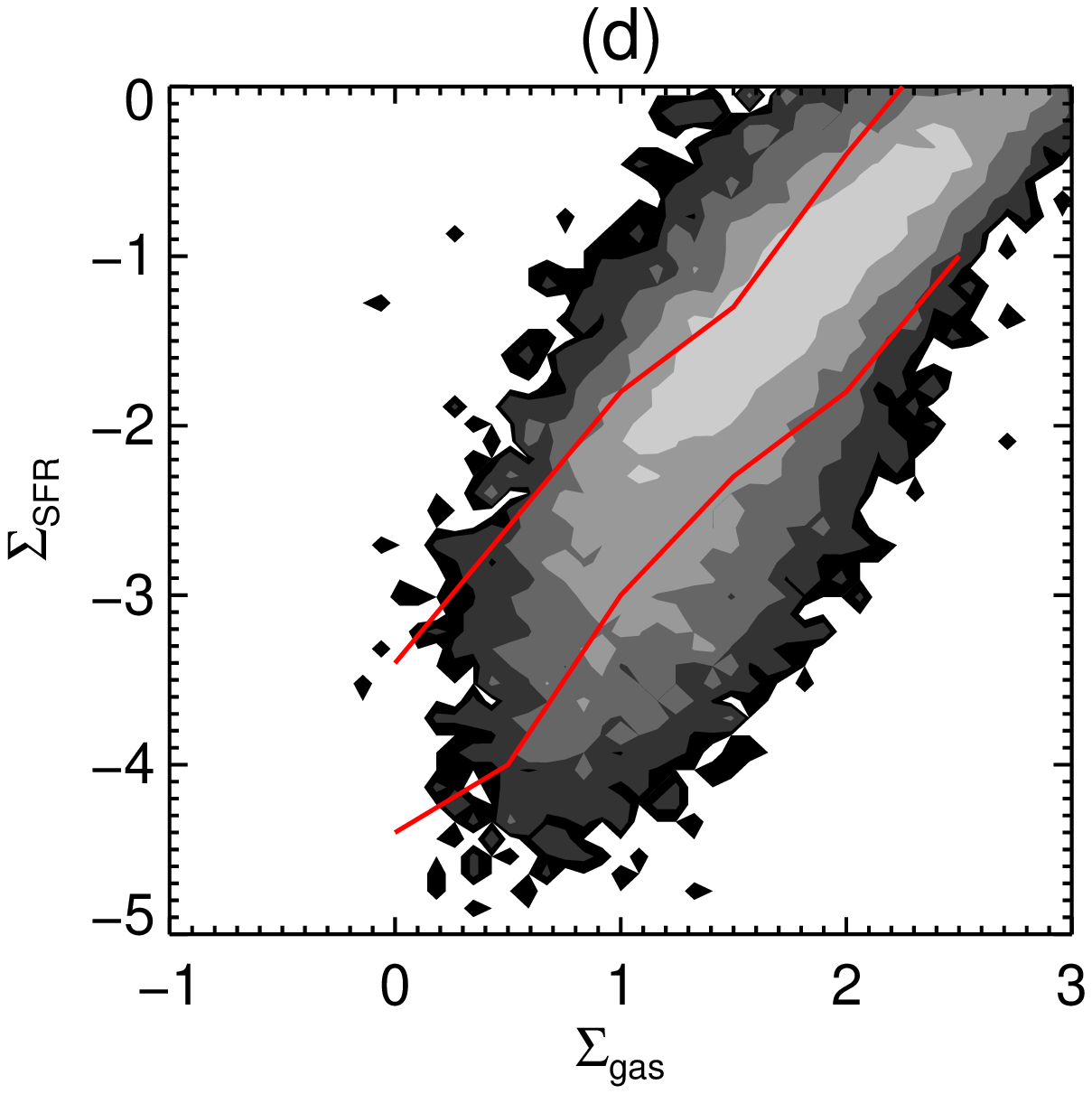}
\caption{Observational parameters derived from the SDSS sample. a. Comparison of stellar mass derived from this work [M17] vs. that derived from \citep{kauffmann03} [K03] (modified to a Salpeter IMF for consistency). For an assumed unity slope the rms deviation is 0.16 magnitudes. A linear fit shows a slightly lower slope (0.91) for M17 with respect to K03. The rms deviation from this line is 0.07 magnitudes. The origin of this slight difference of slope is beyond the scope of this paper and has no impact on the preliminary results we present. b. Comparison of SFR derived from this work to that derived by \cite{salim07} from GALEX UV. Agreement is good with rms deviation 0.2 dex. c. SFR vs M(gas) from observed sample. SFR shows a steep dependence on M(gas) for log M(gas) $>$ 9 and even steeper at lower mass. d. SFR density ($\Sigma_{SFR}$ in $M_{\odot} yr^{-1} kpc^{-2}$ vs gas surface density $\Sigma_{gas}$ in $M_{\odot} pc^{-2}$. Red lines show approximate range of observations from \cite{bigiel08} and \cite{wyder09}.
\label{fig_masscomp}}
\end{figure}

We compare the stellar mass derived by our new approach to that derived by \cite{kauffmann03} in Figure \ref{fig_masscomp}a. The derived masses compare well, with an rms deviation of $\sim$ 0.16 dex around a unity slope. We also find some evidence for a slightly different masses at low and high values as indicated by a best-fit slope of 0.9 for the comparison. This small difference does not affect our preliminary findings discussed below. We compare in Figure \ref{fig_masscomp}b our SFR to the SFR derived by \cite{salim07} also from GALEX UV and an otherwise independent method. The agreement is good with rms deviation of 0.22 dex (comparable to the scatter found by \cite{salim07} of 0.17). We show the SFR vs. M(gas) that we derive. in Figure \ref{fig_masscomp}c. Finally, we show the globally averaged SFR density vs. gas mass density (the standard Schmidt-Kennicutt law \cite{kennicutt98}), compared to results from \cite{bigiel08} for local galaxies.

\subsection{Application 1: Quenching and Starbursts in the Green Valley}\label{sec_app1}

One of our main goals with this technique is to understand the transition of galaxies between the star-forming, blue sequence (or ``main sequence'')  and the passively evolving red sequence. In previous papers \citep{martin07,goncalves12} we have evaluated the timescales required for a galaxy to quench star formation and complete the transition from blue to red, both at low \citep[$z\sim 0.1$;][]{martin07} and intermediate \citep[$z\sim 0.8$;][]{goncalves12} redshifts, using a combination of the $NUV-r$ color and the spectroscopic indices \dn and \hda. Nevertheless, those papers assume a simplistic model of star formation histories in which galaxies move single-handedly from blue to red sequence with exponentially declining star formation rates. We do know, however, that some intermediate-color galaxies are actually {\it bursting}, getting temporarily bluer perhaps due to a sudden inflow of gas and subsequent star formation episode \citep[e.g.,][]{rampazzo07,thomas10,thilker10,salim12,fang12}.

Recognizing this two-way flow, the Star Formation Acceleration (SFA) is an appropriate measure of the rate of color evolution across the Green Valley. Again, SFA is positive for quenching galaxies, and negative for galaxies undergoing starbursts. Figure \ref{fig_samplefits} shows the result for SFA for model galaxies and Figure \ref{fig_influence} shows the observable influence function.

\begin{figure}
\plotone{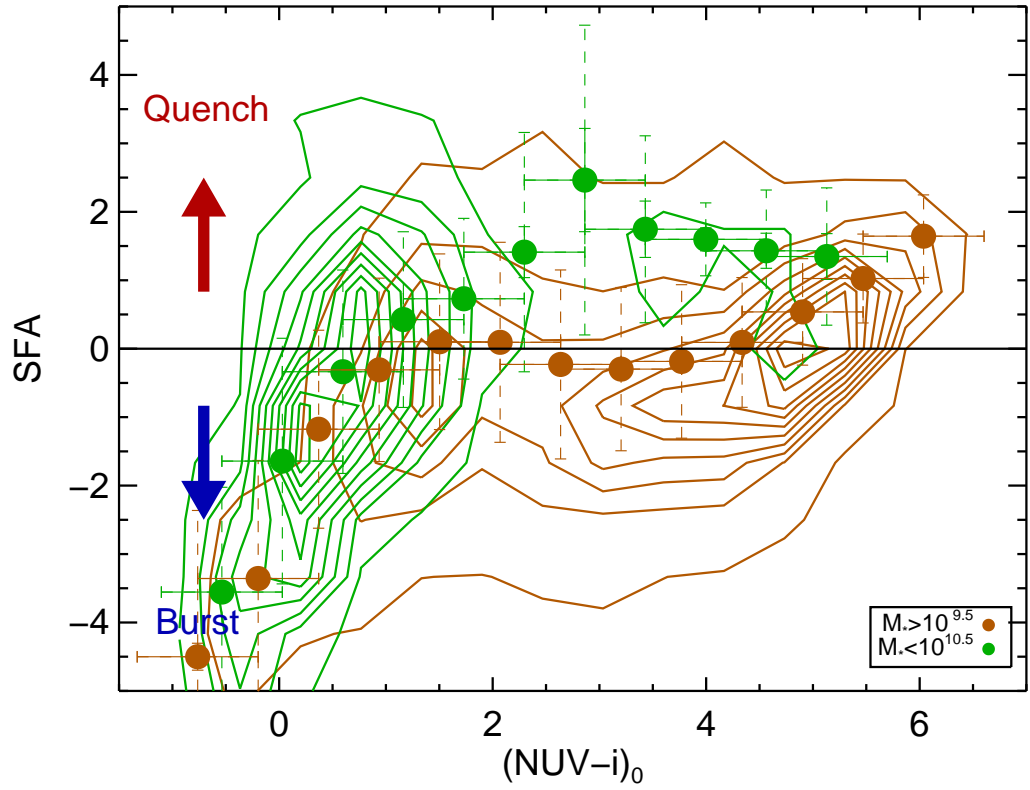}
\caption{Star Formation Acceleration (SFA) vs. (NUV-i)$_0$ for SDSS galaxies in two mass bins cut at transition mass $M<10^{10} M_\odot$ and $M>10^{11} M_\odot$. Contours show distribution of galaxies in the two mass bins. Dots and error bars show mean SFA and error in color bins. The following trends are apparent. Bluer galaxies of both blue and red sequences have bursting SFAs, while redder galaxies of both sequences tend toward quenching SFA. Also, at all (NUV-i)$_0$ colors, on average, lower mass galaxies have higher SFA (more quenching) than higher mass galaxies. 
\label{fig_sfa_nuvi}}
\end{figure}

We applied this to the identical set of galaxies used in \cite{martin07}, and Figure \ref{fig_sfa_nuvi} shows the resulting SFA vs. extinction-corrected NUV-i color in two mass bins. Several phenomena can be seen in this figure. Ignoring mass-dependence for the moment, blue-sequence galaxies show colors correlated with their SFA -- the bluest galaxies have negative, ``bursting'' SFAs, while redder blue-sequence galaxies are ``quenching''. The red sequence has a similar ``tilt'' in the diagram: the bluest galaxies have negative, ``bursting'' SFAs, while redder red-sequence galaxies are ``quenching''. The origin of some of the spread in both sequences can be ascribed to recent changes in the SFR. 

We can plot the color derivatives on the sSFR vs. stellar mass diagram. We show this in Figure \ref{fig_sfa_flux}. This diagram represents a first attempt to capture the ``flow'' of galaxies on the color-magnitude diagram (or equivalent sSFR-mass diagram). In this diagram red arrows represent average quenching and blue average bursting for galaxies in each sSFR-mass bin. The total length of the two arrows is proportional to the rms spread of the SFA, while the relative proportion of red and blue depends on the mean SFA (see caption). The head of each arrow corresponds to the current mass-sSFR, while the tail is the previous location on the diagram scaled to roughly 100 Myr in the past. We can also calculate the mean SFA in each sSFR-mass bin. This is shown in Figure \ref{fig_sfa_ssfr_mstar}.

\begin{figure}
\plotone{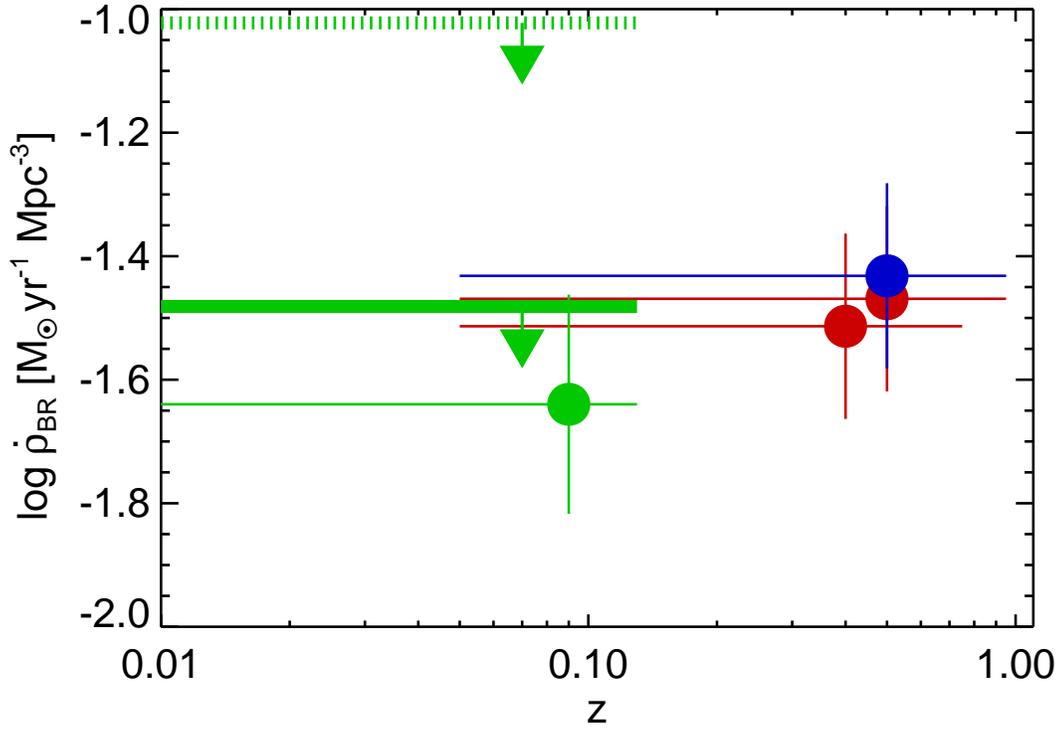}
\caption{Total mass flux across the Green Valley (green dot with error bars) estimated using SFA and the galaxy sample of M07. 
Green dashed shows result of M07 method 1 (no extinction correction, color derivative calculated for the mean of all galaxies in the color-magnitude bin based on monotonically quenching star formation histores), green solid shows
result of M07 method 3 (extinction correction, color derivative calculated for each galaxy based on monotonically quenching star formation histories). Both were interpreted in M07 as upper limits because of the possible presence of bursting galaxies. Red points show mass flux estimated from
red sequence evolution of \cite{faber07}. Higher point is based on
evolution over $0<z<1$, while lower point on $0<z<0.8$. Blue point
shows estimate $-\dot{\rho}_B$ based on blue sequence evolution (derived in M07). 
\label{fig_rhobr}}
\end{figure}

In M07 we reported a measurement of the mass flux of galaxies across the green valley as an upper limit, because we used a simple monotonic quenching model to derive the color-derivative (dy/dt, now relabeled SFA). Using the same sample but revising the color derivative in each mass bin, we can calculate the true mass flux from blue to red taking into account net bursting and quenching. The revised flux vs. mass is given in Table \ref{tab_method4}. Our new mass flux (calling this method 4 to maintain continuity with the three methods presented in M07) is $\dot{\rho}_{BR}=(2.3\pm0.07)\times 10^{-2} M_\odot yr^{-1} Mpc^{-3}$. It is entirely consistent with the value derived by M07, and also with the estimates based on the mass evolution of the blue sequence \citep{blanton06,martin07} and red sequence \citep{faber07}. We plot this result in Figure \ref{fig_rhobr}.

\begin{figure}
\plotone{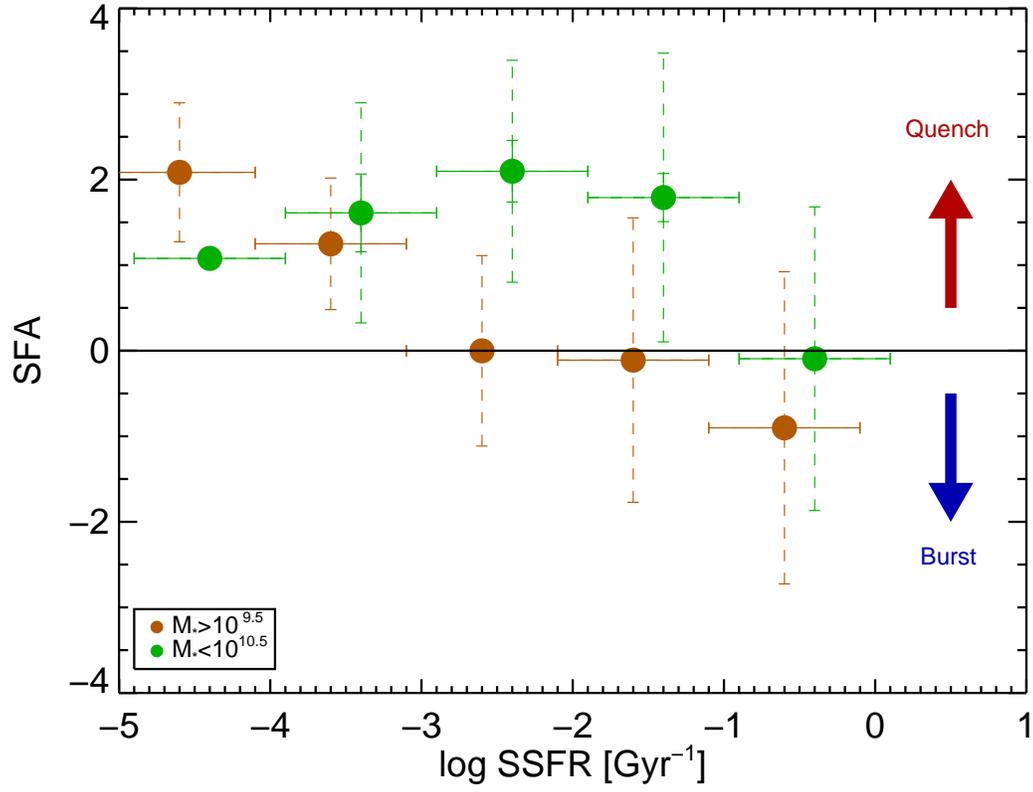}
\caption{Average Star Formation Acceleration (SFA) in several sSFR bins for SDSS galaxies in two mass bins cut at transition mass $M_c=10^{10} M_\odot$. As in Figure \ref{fig_sfa_nuvi}, the following trends are apparent. High sSFR galaxies have bursting SFAs, while low sSFR galaxies tend toward quenching SFA. For log(sSFR)$<-1.5$ Gyr$^{-1}$ lower mass galaxies have higher SFA (more quenching) than higher mass galaxies, which average zero quenching (equal numbers of quenching and bursting galaxies). Dashed lines show the full range of the distribution, while solid vertical error bars show the resulting standard error of the mean SFA.
\label{fig_sfa_ssfr}}
\end{figure}

Now consider the dependence on stellar mass. Lower mass galaxies in the green valley are mostly quenching, while higher mass galaxies are both quenching and bursting. This is demonstrated in the sSFR-mass diagrams Figure \ref{fig_sfa_flux} and Figure \ref{fig_sfa_ssfr}. In Figure \ref{fig_sfa_ssfr} we have calculated average SFA and display them vs. specific SFR (sSFR) for two mass cuts. The mean SFA is 1-3 higher for galaxies with $M_* < 10^{9.5} M_\odot$ compared to galaxies with $M_* > 10^{10.5} M_\odot$. A plausible scenario for this is given in Figure \ref{fig_sfa_model}: lower mass galaxies are accreting and becoming satellite galaxies, having their star forming gas tidally and/or ram-pressure stripped, while higher mass galaxies are receiving this gas and reacting with new star formation. These mass differences are extremely important for galaxy models, and obtaining significant numbers of low mass green-valley galaxies and comparing them to high mass galaxies requires an analysis of a larger SDSS/GALEX dataset.  

It is interesting to compare these observed results to the predictions of the semi-analytic models used to generate the star formation histories and parameter coefficients. As we discuss in the appendix (\S \ref{sec_model_comp}), the models predict trends that are qualitatively similar but quantitively much weaker than those we observe. The observed results are quite distinct from the model predictions. 

\subsection{Application 2: The AGN/SFA connection}

\begin{figure}
\plotone{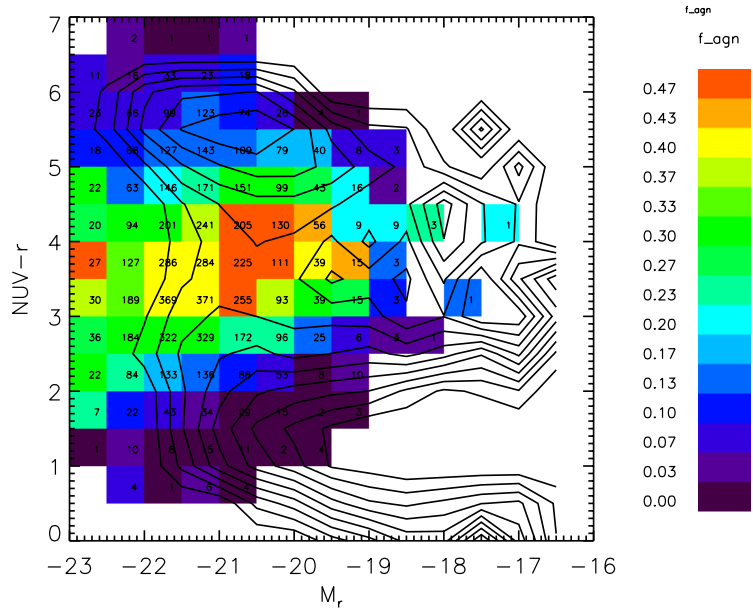}
\caption{Color-magnitude diagram  (contours in NUV-r vs. M$_r$, extinction corrected) from Martin et al., (2007a). AGN fraction in each color-magnitude bin shows that AGNs mostly occupy the green valley.
\label{fig_cmd_agn}}
\end{figure}

\begin{figure}
\plotone{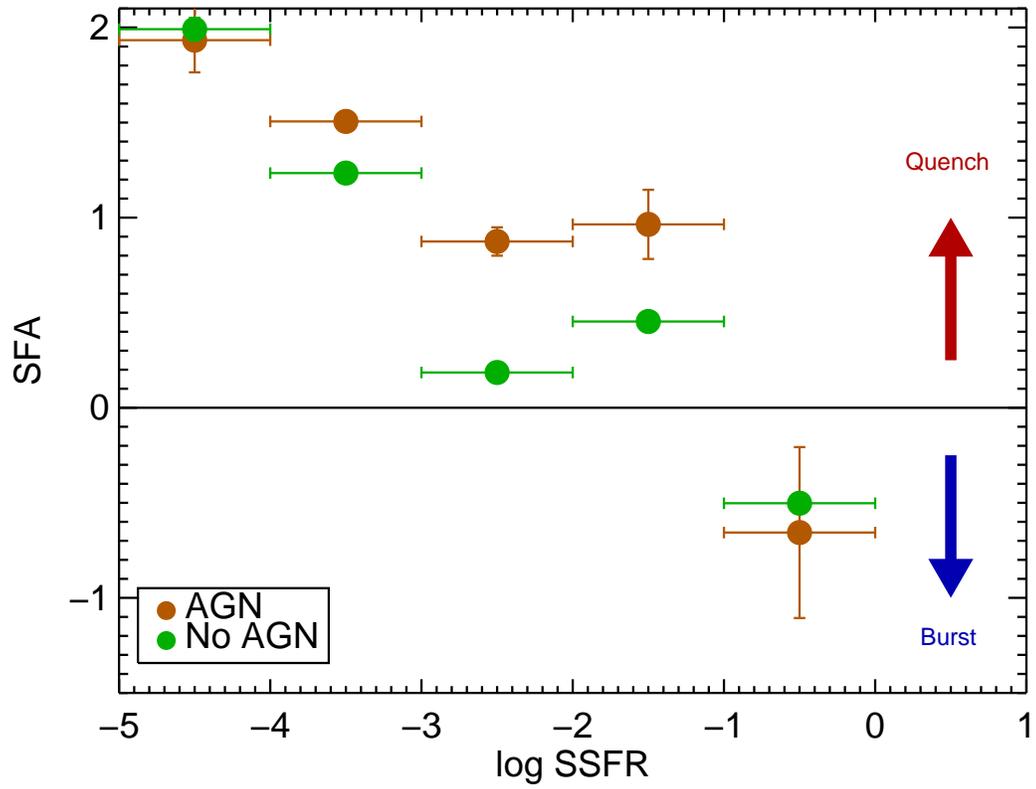}
\caption{Average Star Formation Acceleration (SFA) in several sSFR bins for SDSS galaxies for AGNs and non-AGNs. As in Figure \ref{fig_sfa_ssfr}, SFA increases at lower sSFR. Since low mass galaxies dominate the number density, the average is net quenching at low sSFR. Galaxies with AGN show detectably higher SFA (quenching) than galaxies without AGN. Vertical error bars show the standard error of the mean SFA.
\label{fig_sfa_agn}}
\end{figure}

AGNs are potentially powerful source of feedback that could accelerate quenching and maintain galaxies on the red sequence \citep{croton06,martin07,nandra07,schawinski09}. Furthermore, there is growing evidence that quenching (especially at high stellar masses) might be related to the growth of stellar density in the central of the galaxy, probably due to AGN activity and concomitant bulge growth \citep[e.g.,][]{cheung12,mancini15}.
As Figure \ref{fig_cmd_agn} shows, AGNs preferentially occupy the green valley. We would like to attempt to answer a simple physical question: All else being equal, does the presence of an AGN accelerate quenching in transition galaxies? There is preliminary evidence for this, which we show in Figure \ref{fig_sfa_agn}. At intermediate sSFR, the presence of an AGN appears to accelerate quenching by roughly a factor of 2-3. This would appear to support a scenario in which the presence of an AGN might also be connected with a starburst event \citep[e.g.,][]{king05,gaibler12,rovilos12}, and only unequivocally quenches star formation at later stages, when feedback drives the gas away \citep{springel05,dimatteo05}.

However, a large, statistically robust sample is required to confirm this tentative conclusion. AGN fraction correlates with many other properties, and it must be established that these correlations do not artificially create this dependence. The larger GALEX Legacy Survey/SDSS sample will allow us to test this dependence while other correlates are held fixed, and even investigate whether there is a relation between quenching timescales and AGN luminosities. One of the future goals of this study is to firmly establish whether AGNs accelerate quenching, and under what circumstances.

Our preliminary results can also be used to place some constraints on the formation of the most massive ($M_* > 10^{11.5} M_\odot$) quiescent galaxies. Let's use a cut of log SSFR(Gyr$^{-1}$) $=$ $-2$ to separate star-forming and quiescent systems (used in the literature and is evident from the distribution of galaxies in Figure \ref{fig_sfa_flux}). If the most massive ($M_* > 10^{11.5} M_\odot$) quiescent galaxies are the result of dry mergers between already quiescent less massive systems, then in principle, there should not be a change in their SFA. However, we clearly see in Figure \ref{fig_sfa_ssfr_mstar} that even the most massive quiescent systems show some degree of bursting. Wet mergers can qualitatively explain the bursting phase for them. The most bursting is happening in the most massive star-forming systems as seen in Figures \ref{fig_sfa_ssfr_mstar} and \ref{fig_sfa_ssfr}. These star-forming systems are likely going through wet major mergers that result in gas in the outskirts of them falling toward the center and getting compressed, causing the burst of star-formation. Very massive star-forming systems ($M_* > 10^{11.5} M_\odot$ and log(SSFR) $>$ -0.5) are rare (see, e.g.; Figure \ref{fig_sfa_ssfr_mstar}) because they have already been quenched and moved to the massive quiescent population likely through wet major mergers of less massive star-forming systems. Therefore, part of the evolution of the most massive quiescent galaxies ($M_* > 10^{11.5} M_\odot$) is due to wet major mergers of less massive star-forming systems. We note that wet minor mergers can have a similar effect too, without changing the mass of the massive quiescent galaxies much. Star Formation Jerk (SFJ) and a larger sample can potentially help distinguish between these scenarios. 

Interestingly, wet major mergers might also explain what we see in Figure \ref{fig_sfa_agn} for AGNs. Wet major mergers tend to rejuvenate the nuclear activity but with some time delay after the star-bursting phase (due to star formation). According to Figure \ref{fig_sfa_agn}, for high SSFR values (star-forming phase), both AGN and non-AGN hosts are bursting (in the star-formation phase of merger) but after a while, they enter the quenching phase with AGN hosts showing higher quenching possibly due to the revived nucleus (as mergers cause the gas to funnel toward the nucleus), which is subsequently followed by outflows/feedback to help quench galaxies more effectively. SFJ contains information about the timescale of quenching/bursting events and can potentially be used to constrain this picture. In a following paper, we will study this in more details.

\section{Discussion and Summary} \label{sec_discussion}

\subsection{Issues and Caveats}

Aperture and Volume Effects SDSS spectroscopy is obtained with 3 arcsecond fibers which often do not subsume the full galaxy. M07 discussed this effect and dismissed it as not significant, mainly on the strength of no detected average redshift dependence. There are small variations in $<SFA>$ vs. redshift that may be correlated with large scale structure. There is no trend with increasing redshift. The mass trend of SFA does not diminish when the redshift range is restricted to $0.05<z<0.10$. This indicates that neither aperture, color selection, mass selection, or volume effects explain the mass trend. 

{\it Extinction} We considered a number of variants of the extinction law behavior to determine whether our approach impacted the star formation history extraction. In all cases the derived extinction has sensible dependence on SFR, SSFR, metallicity and gas mass. 
As we noted above, even when the extinction law is permitted to vary randomly between Milky Way and Calzetti, the rms error in the A$_{FUV}$ rises only to $\sim 0.5$ magnitudes. Other than making subtle changes in the distribution of galaxies in the Mass-SFR and Mass-SSFR diagram, the details of the extinction correction do not significantly impact SFA. We defer to a future paper a comparison of the extinction correction with direct methods that use the MIR/FIR and FUV/NUV luminosity. 

{\it Model Biases} It is important to ascertain whether the particular SAMs we have chosen to generate star formation histories are biasing the results for the observed SDSS sample. As we mentioned earlier, we believe that the SAMs provide a space of possible star formation histories, and if those histories span a similar space as actual galaxies (not necessarily with the same demographics), then the SFA we derive will not be sensitive to the models. We show in Appendix \S\ref{sec_model_comp} that the SAMs give a quantitatively different SFA vs. mass and sSFR than the observed galaxies.

\begin{figure}
\plotone{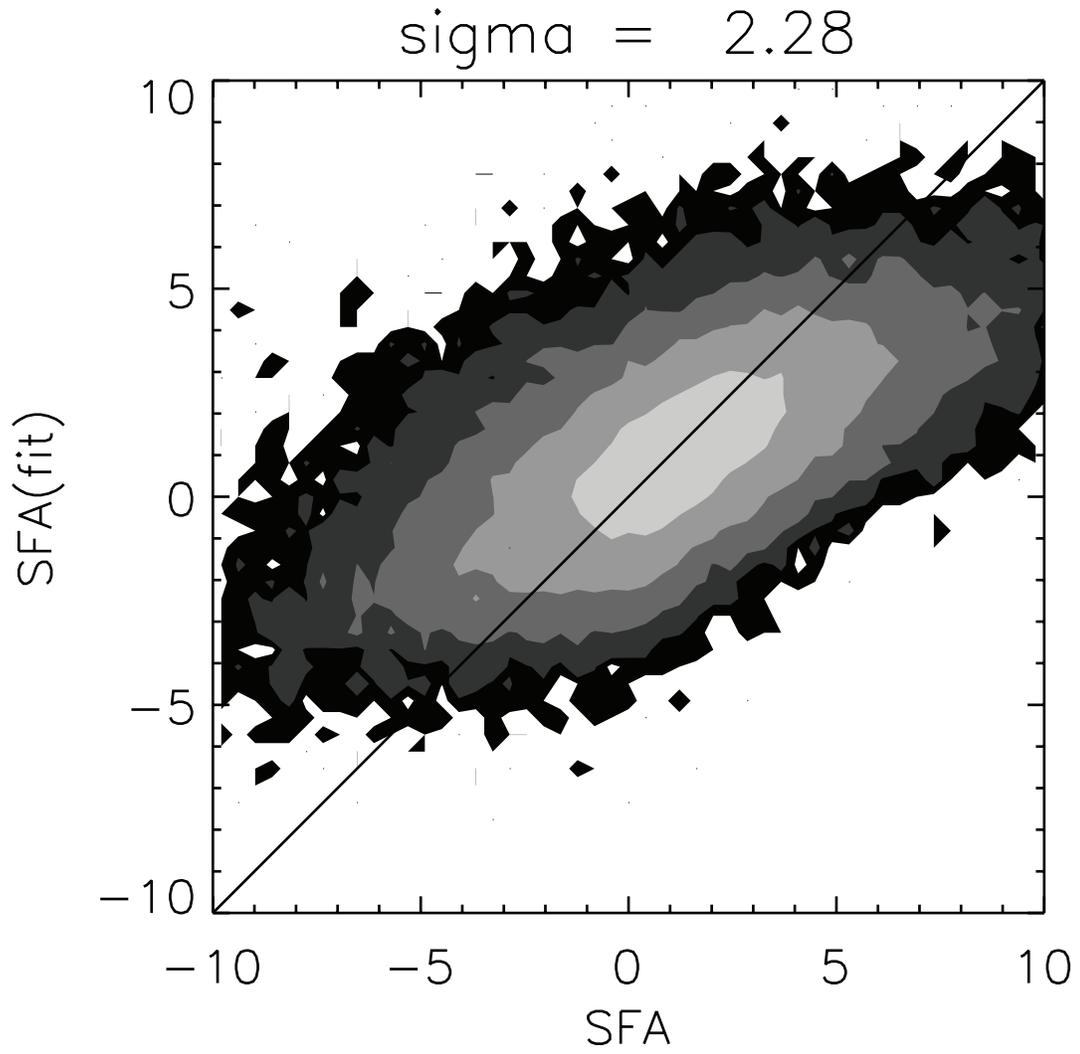}
\caption{Recovered SFA vs model SFA in test sample for which a largr random SFR component as been added to decouple the star formation history from other physical parameters that affect observables such as extinction (compare to Figure 6f). Ratio of fitting error ($\sigma=2.28$) relative to spread of SFA is approximately the same as in the reference models.
\label{fig_sfa_random}}
\end{figure}

We experimented with changing the star formation histories in the SAMs by adding a large random component (by replacing SFR with 2*SFR*r where $0<r<1$ is a uniform random deviate). The purpose of this was to show that the SFA recovery is tied to star formation history alone and not some other observable quantity (such as extinction) given by the models.   Even here we still recover the relationship between observables and SFA with similar coefficients and a similar ratio of fit noise to the total dispersion in SFA (which in this case is larger because of the very noisy star formation histories). This fit is shown in Figure \ref{fig_sfa_random}. This occurs is in spite of the large decoupling between the star formation histories and the other physical parameters in the models with a random star formation history. Just as SFR  (the derivative of stellar mass) is traced by FUV, NUV, or H$\alpha$ (extinction corrected) in model independent way, so does SFA (effectively the derivative of log sSFR) traces the color derivative in an essentially model-independent way. The caveat to this discussion is metallicity, which we turn to next.

{\it Metallicity} Spectral indices and photometric colors are dependent on the metallicity of the stars producing them as well as on the star formation histories. As we discussed in \S\ref{sec_pop_synthesis}, model metallicities are incorporated following the SAM metallicity evolution for each galaxy. Thus to first order metallicity effects are accounted for, to the extent that the model galaxy metallicity evolution matches that of the observed galaxies. We have checked to see whether uncorrected metallicity variations in the \hda-\dn relation can produce the mass trends that we observe. Consider a SSFR range of $-3<log SSFR<-2$, in three mass bins ($9<log M_*<10$,  $10<log M_*<11$, and $11<log M_*<12$). These give mean \dn of 1.54, 1.70, and 1.76. Using the mass-metallicity relation of \cite{tremonti04} and the \hda-metallicity variation for fixed \dn from \cite{bruzual03}, we can calculate $d$\hda$/d(log M_*)$, and using the fitting coefficients $dSFA/d$\hda we find a spurious slope of $dSFA/d(log M_*)=-0.5$ between $9<log M_*<10$ and $10<log M_*<11$, and $dSFA/d(log M_*)=+0.15$ between $10<log M_*<11$ and $11<log M_*<12$. This should be compared with the observed  $dSFA/d(log M_*)=-1.2$ for both cases. Thus even uncorrected metallicity effects in the spectral indices cannot reproduce the observed mass trends. 

\subsection{Summary}

We propose a novel methodology to investigate galaxy properties through use of a combination of photometric and spectroscopic measurements. By using stellar population synthesis models, we are able to recover a large array of physical properties of model galaxies using such combination. In particular, we define a new quantity, star formation acceleration (SFA), which traces the instantaneous time derivative of the specific star formation rate of an individual galaxy by measuring the NUV-i color time derivative, and which is also recovered by use of the aforementioned measurements.

The approach offers the following benefits: 

\begin{enumerate}
\item Physical parameters are derived not by fitting but
by a single matrix of linear coefficients;
\item The method makes no assumptions about star formation histories;
\item Moments of star formation history (the star formation rate and higher derivatives) can be derived non-parametrically; 
\item The method works over all stellar masses with a single set of matrices; 
\item Degeneracies between the derived physical parameters and covariance are explicitly
derived;
\item Error propagation is simple; 
\item The influence of each observable on each derived physical parameter can be calculated and
the resulting sensitivities provide useful context for error analysis and observation planning; 
\item The method is easily generalized to
incorporate new observables (e.g., morphological indices, other line indices, emission line fluxes, sersic indices, environmental parameters) and model-generated physical parameters (e.g., bulge-to-disk ratio, galaxy density);
\item The method is linear and therefore stacked spectra (within constant D$_n(4000)$ bins) can be used to derive average physical parameters. For example, galaxies can be stacked in bins, (e.g., extinction-corrected color-magnitude bins), obtaining an average physical parameter for the bin.
\end{enumerate}

\acknowledgments

GALEX (Galaxy Evolution Explorer) is a NASA Small Explorer, launched in April 2003.
We gratefully acknowledge NASA's support for construction, operation,
and science analysis for the GALEX mission,
developed in cooperation with the Centre National d'Etudes Spatiales
of France and the Korean Ministry of 
Science and Technology. Behnam Darvish acknowledges financial support from NASA through the Astrophysics Data Analysis Program (ADAP), grant number NNX12AE20G. We thank the anonymous referee for valuable comments that strengthed the paper.



{\it Facilities:} \facility{GALEX}, \facility{SDSS}

\appendix

\section{Color and Spectral Index Correction and Impact on SFA} \label{sec_color_correct}
 
We have adjusted model colors and spectral indices so that they are similar to those of the observed sample. We do this so that the range of observational parameters used to extract the physical parameters are comparable to the model range. In order to make this comparison as representative as possible, we use a filtered sample of the model galaxies selected to be detected in SDSS and GALEX NUV as a function of their redshift. In other words the color-correction model sample is magnitude-limited in the same way as the observed sample. We note that the entire model sample was used to derive the regression coefficients, not the filtered sample.

For each \dn bin, we compare the distribution of model and observed colors and indices, notably \hda. We have tried using two methods: simple means and maximizing the cross-correlation. These give results typically within $\sim$0.1 in correction values. Our default is the mean method. For convenience we correct observables to model values, noting that this is equivalent to correcting model values to observable distributions (resulting in modified regression offsets), and permits the application of the published regression coefficients to other data sets.

\begin{figure}
\plotone{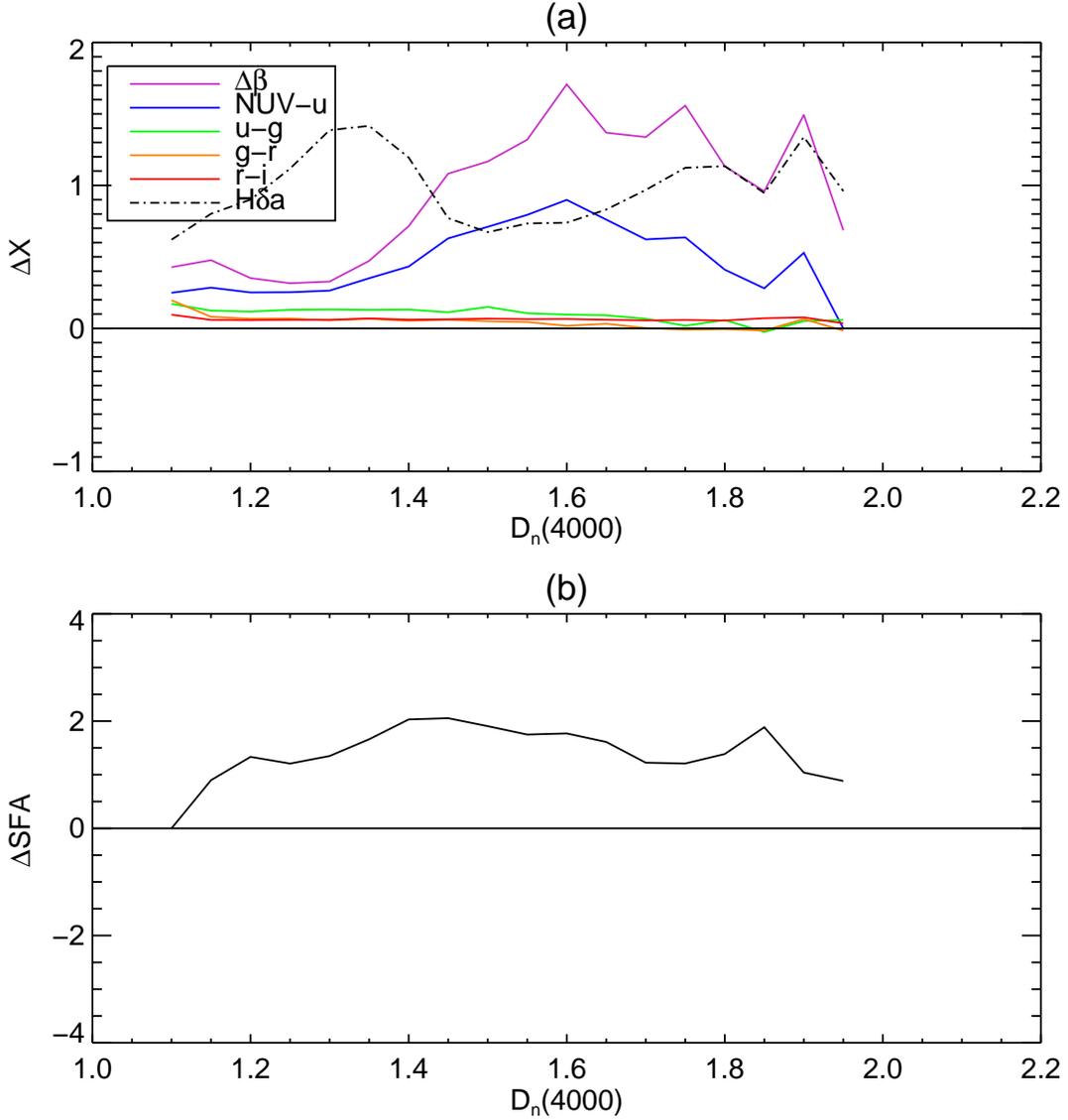}
\caption{a. Changes to colors and \hda vs. \dn required to match observational to model distributions. b. Resulting change to mean SFA produced by color/index changes vs. \dn.
\label{fig_color_correct}}
\end{figure}

We show a summary of the observable color/index changes in Figure \ref{fig_color_correct}a and tabulate them in Table \ref{tab_colcor}. Note that the largest changes are to $\beta$ (the FUV-NUV slope parameter), NUV-u (and correspondingly NUV-i), and \hda. We also show in Figure \ref{fig_color_correct}b the impact on SFA. Over a most of the range of \dn there is an increase in SFA in the range of 0.7-2.0 mag/Gyr, with a mean change of $\Delta SFA=1.35$.

\begin{figure}
\plottwo{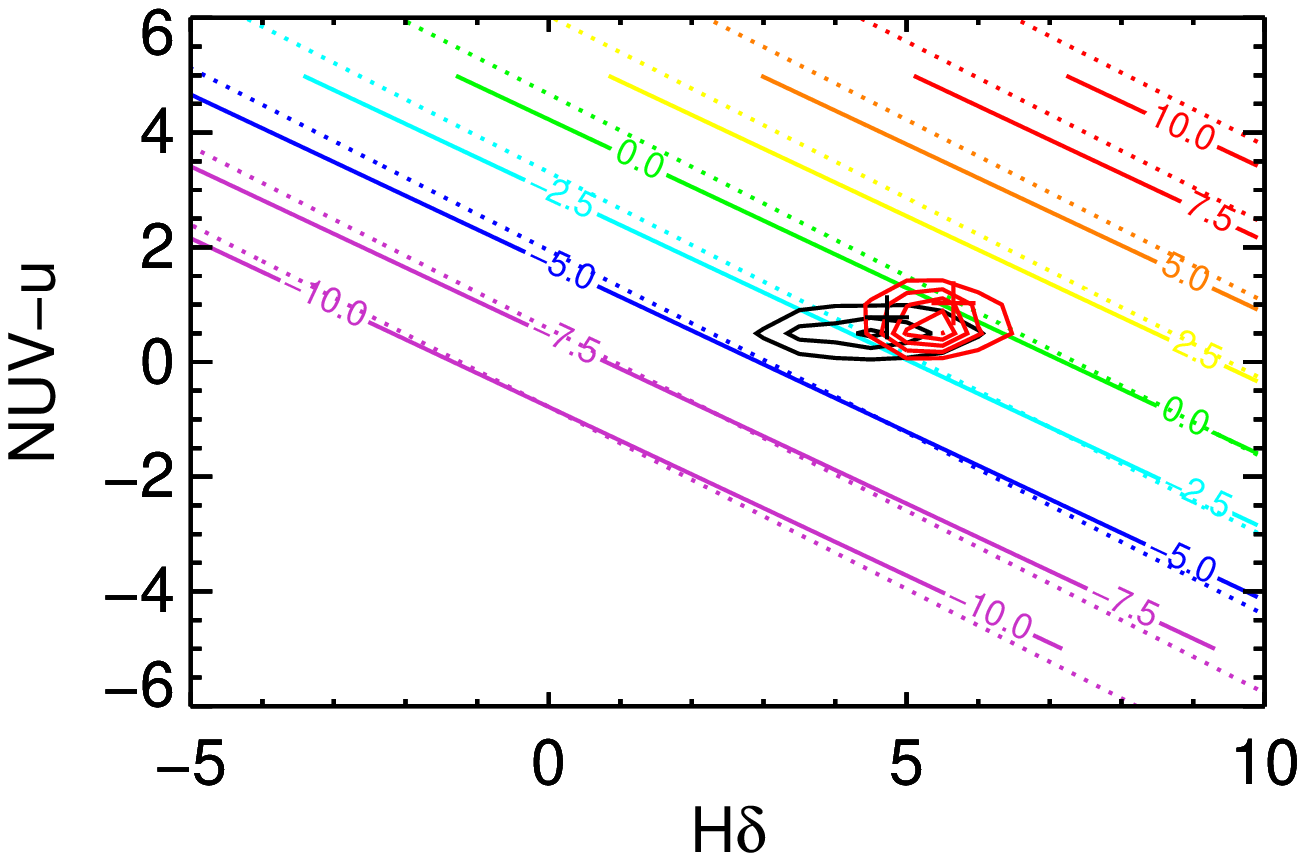}{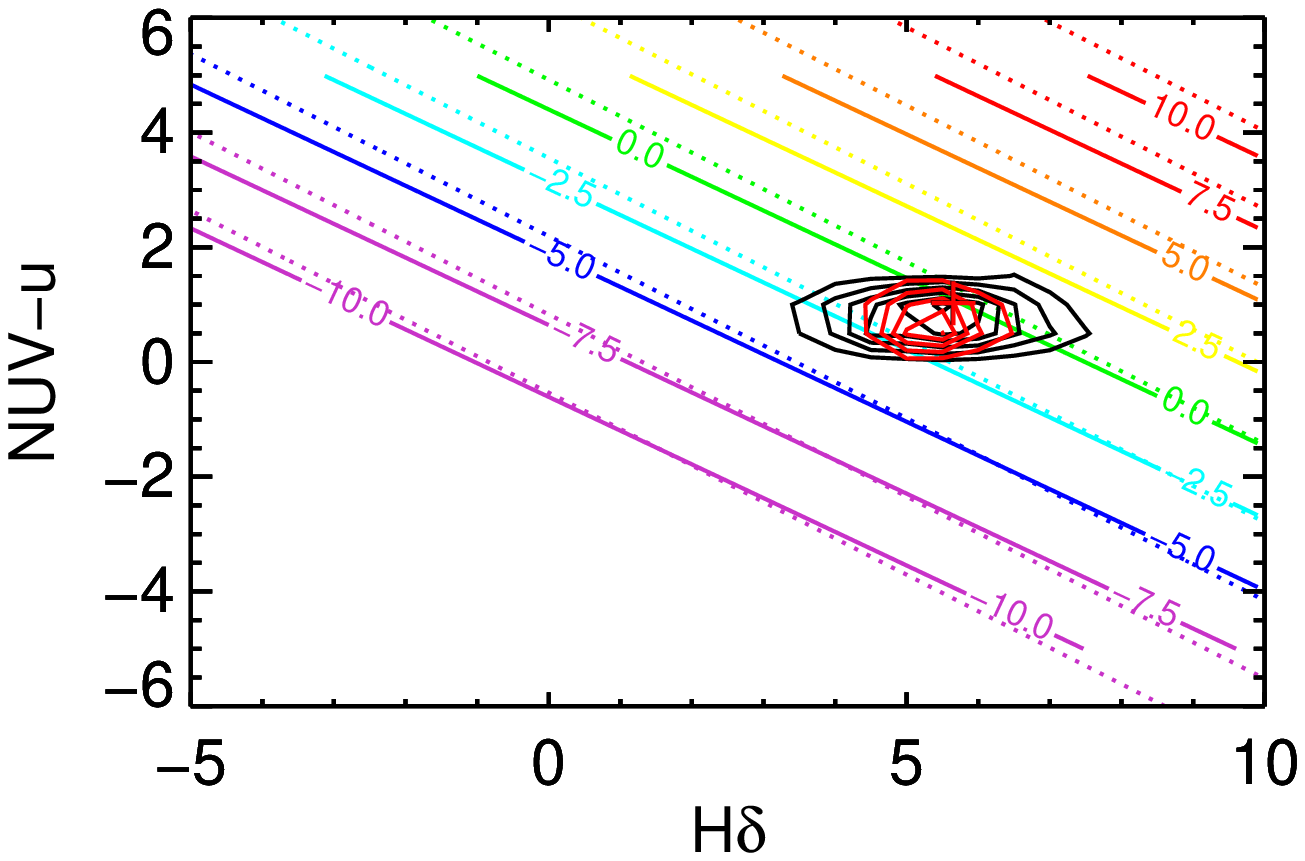}
\caption{a. Distribution of observables (\hda and NUV-u) in black contours, and model values in red contours, with means shown with crosses. For \dn=1.25 bin. Contours show SFA vs. \hda and NUV-u calculated using the mean values of the other observables in this \dn bin. Solid lines shows SFA calculated using FUV-NUV (or $\beta$), while dotted lines show SFA calculated not using FUV-NUV. b. Same as a. with corrected observables and SFA contours calculated using corrected mean observables.
\label{fig_correct_nuvu_1.25}}
\end{figure}

\begin{figure}
\plottwo{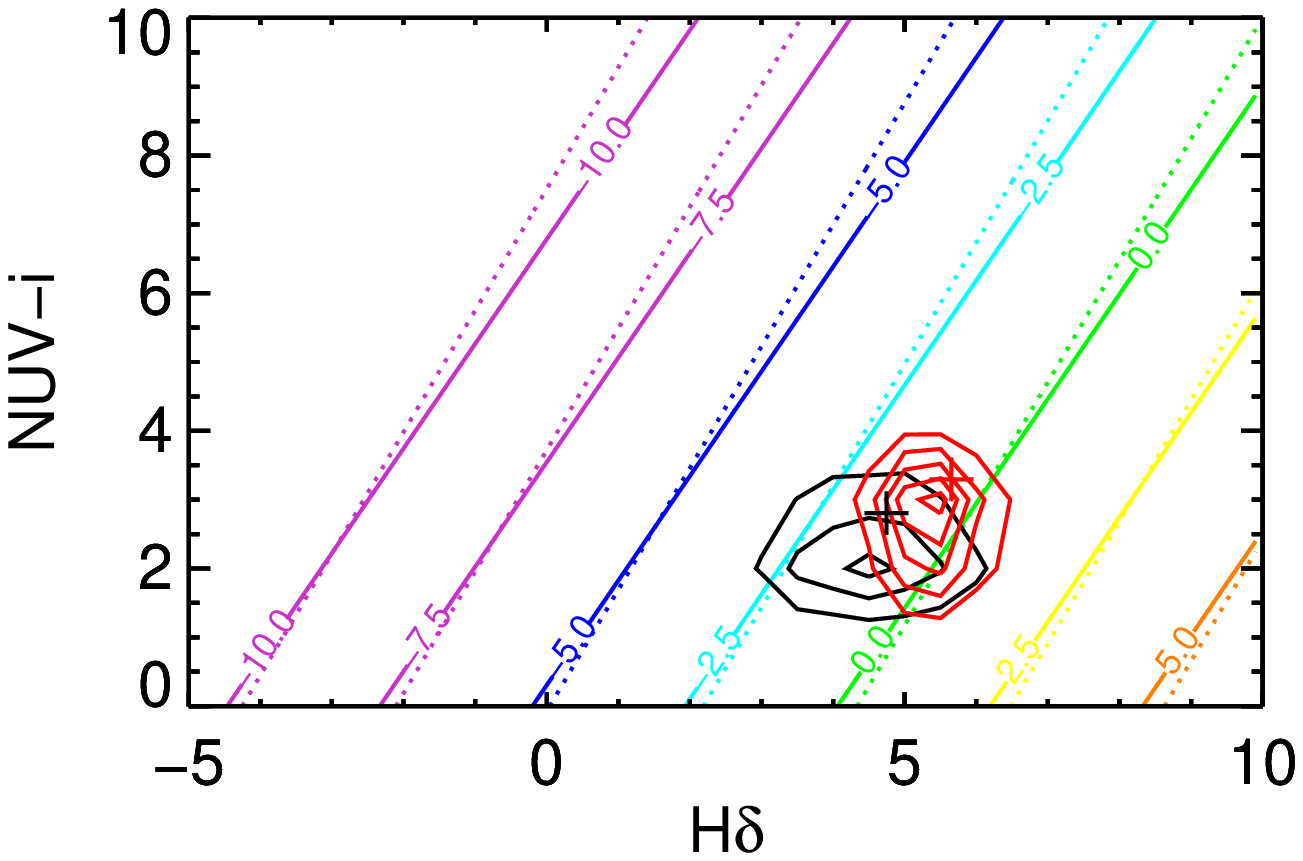}{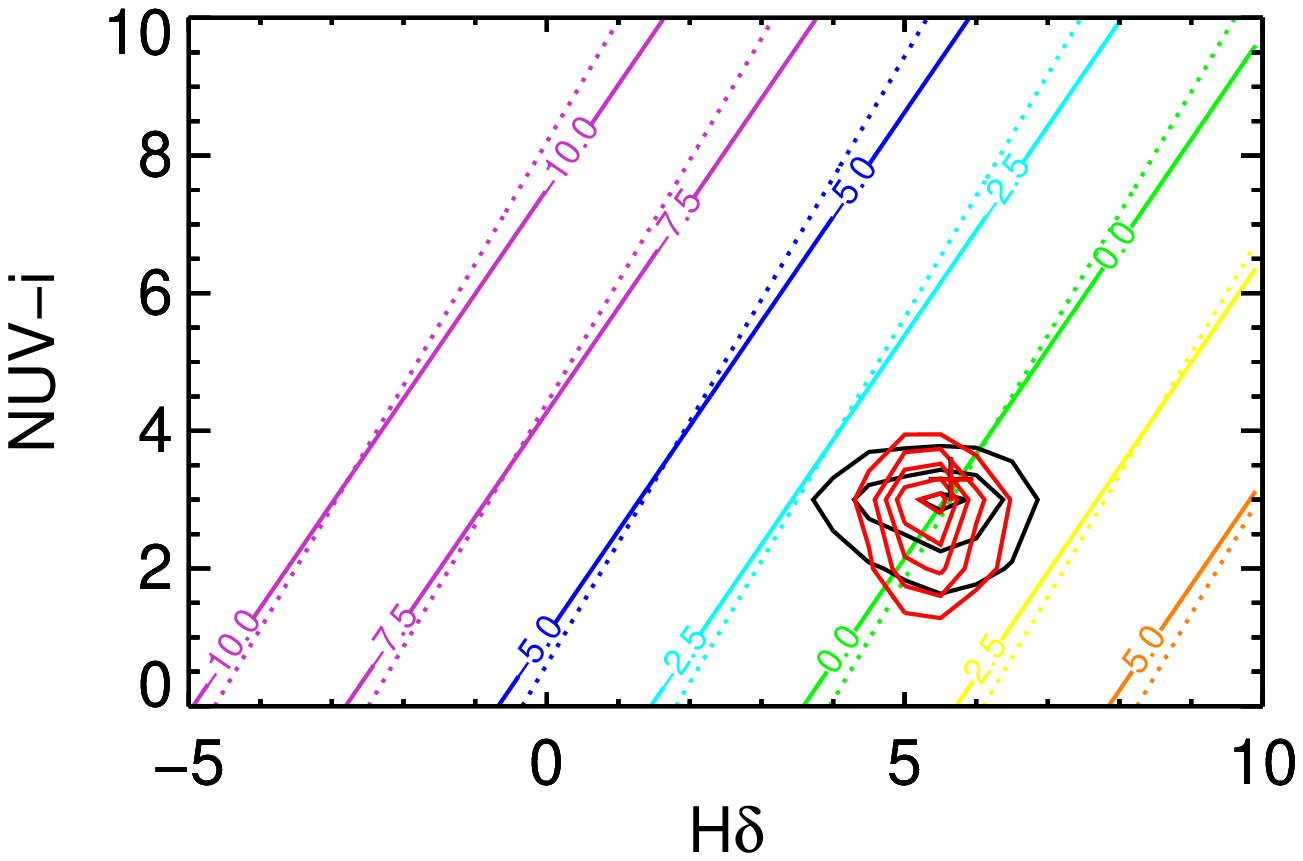}
\caption{a. Same as Figure \ref{fig_correct_nuvu_1.25} for NUV-i vs. \hda.
\label{fig_correct_nuvi_1.25}}
\end{figure}

\begin{figure}
\plotone{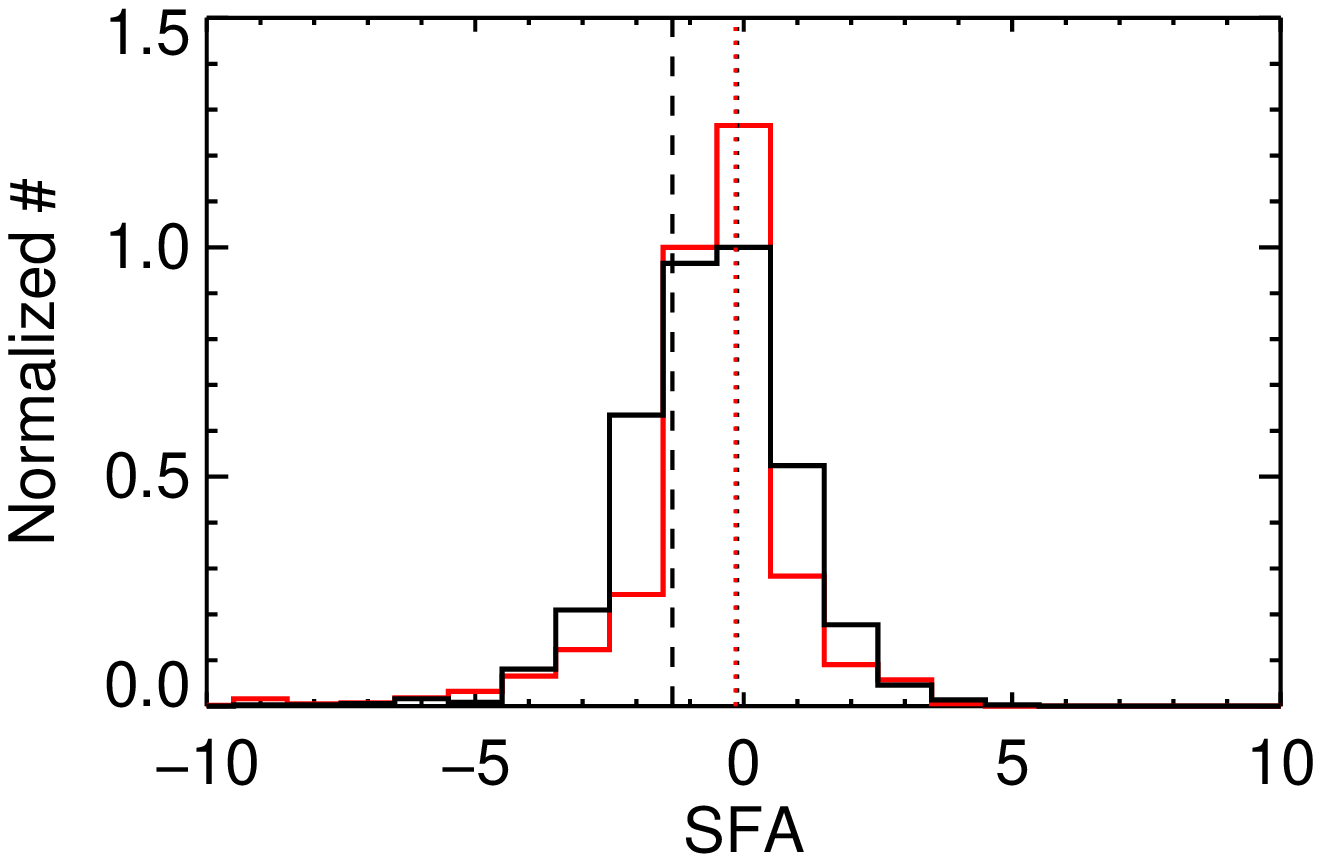}
\caption{{|bf Histogram of derived SFA (with corrected observables, black) vs. model SFA. Dotted lines show mean values (black: observable SFA, red: model SFA). Dashed line shows mean derived SFA for uncorrected observables. All in \dn=1.25 bin.}
\label{fig_hist_sfa_1.25}}
\end{figure}

\begin{figure}
\plottwo{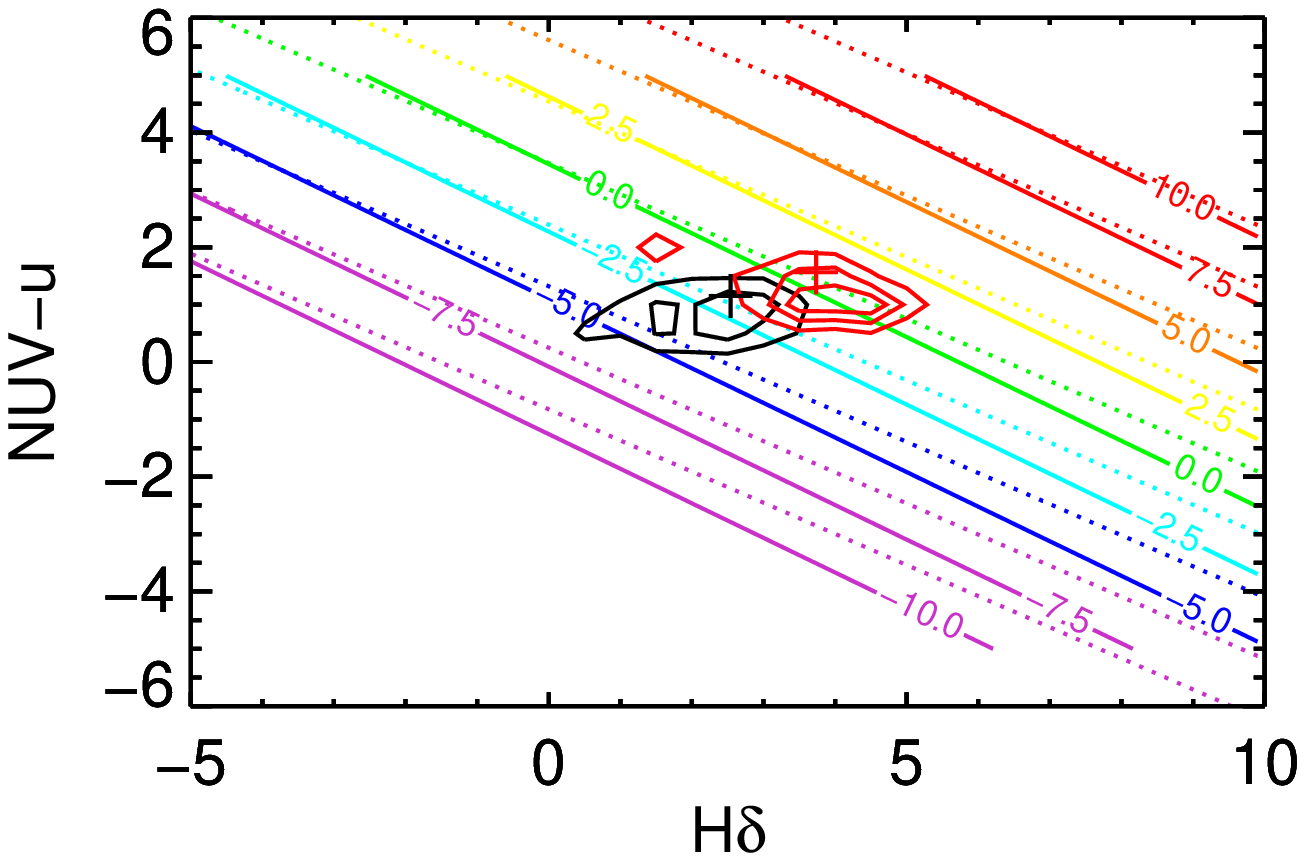}{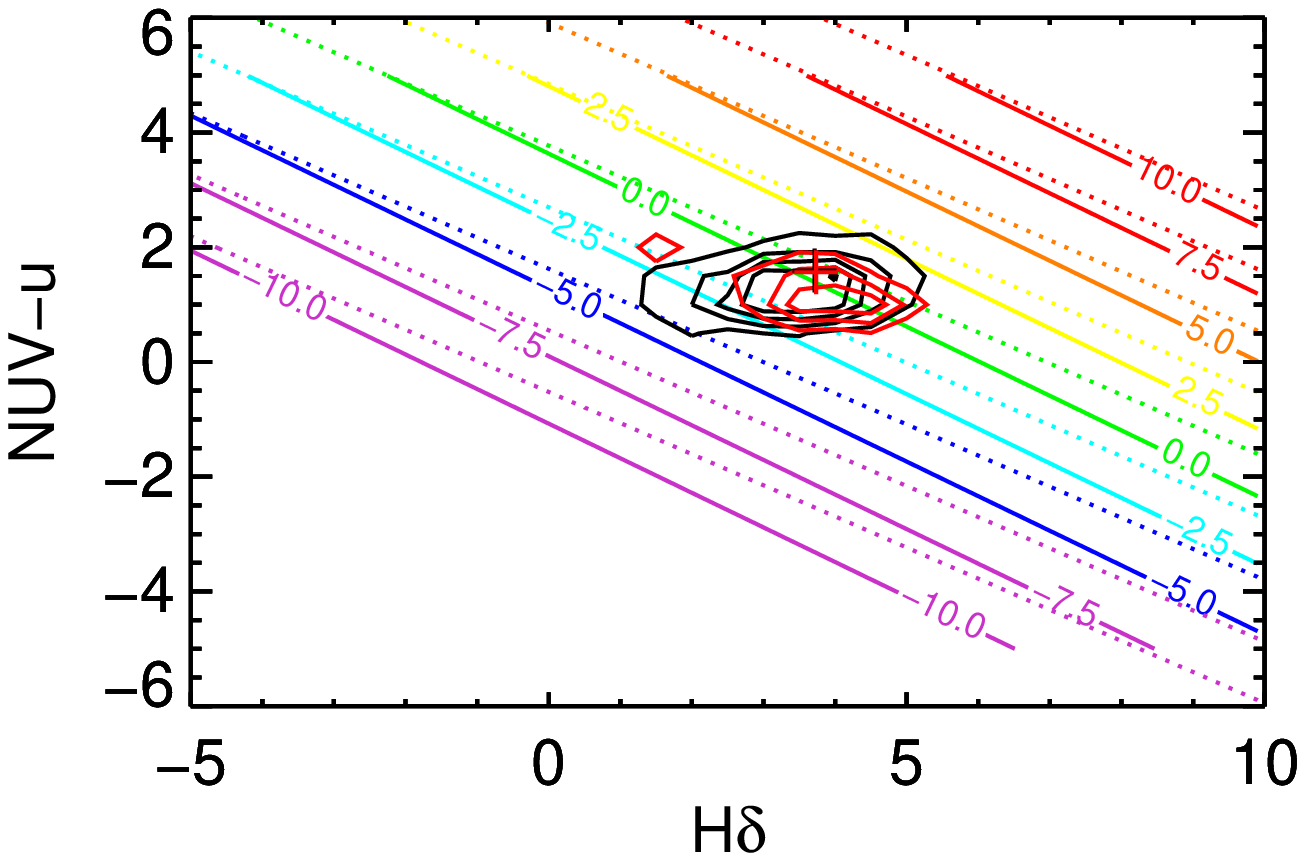}
\caption{a. Distribution of observables (\hda and NUV-u) in black contours, and model values in red contours, with means shown with crosses. For \dn=1.45 bin. Contours show SFA vs. \hda and NUV-u calculated using the mean values of the other observables in this \dn bin. Solid lines shows SFA calculated using FUV-NUV (or $\beta$), while dotted lines show SFA calculated not using FUV-NUV. b. Same as a. with corrected observables and SFA contours calculated using corrected mean observables.
\label{fig_correct_nuvu_1.45}}
\end{figure}

\begin{figure}
\plottwo{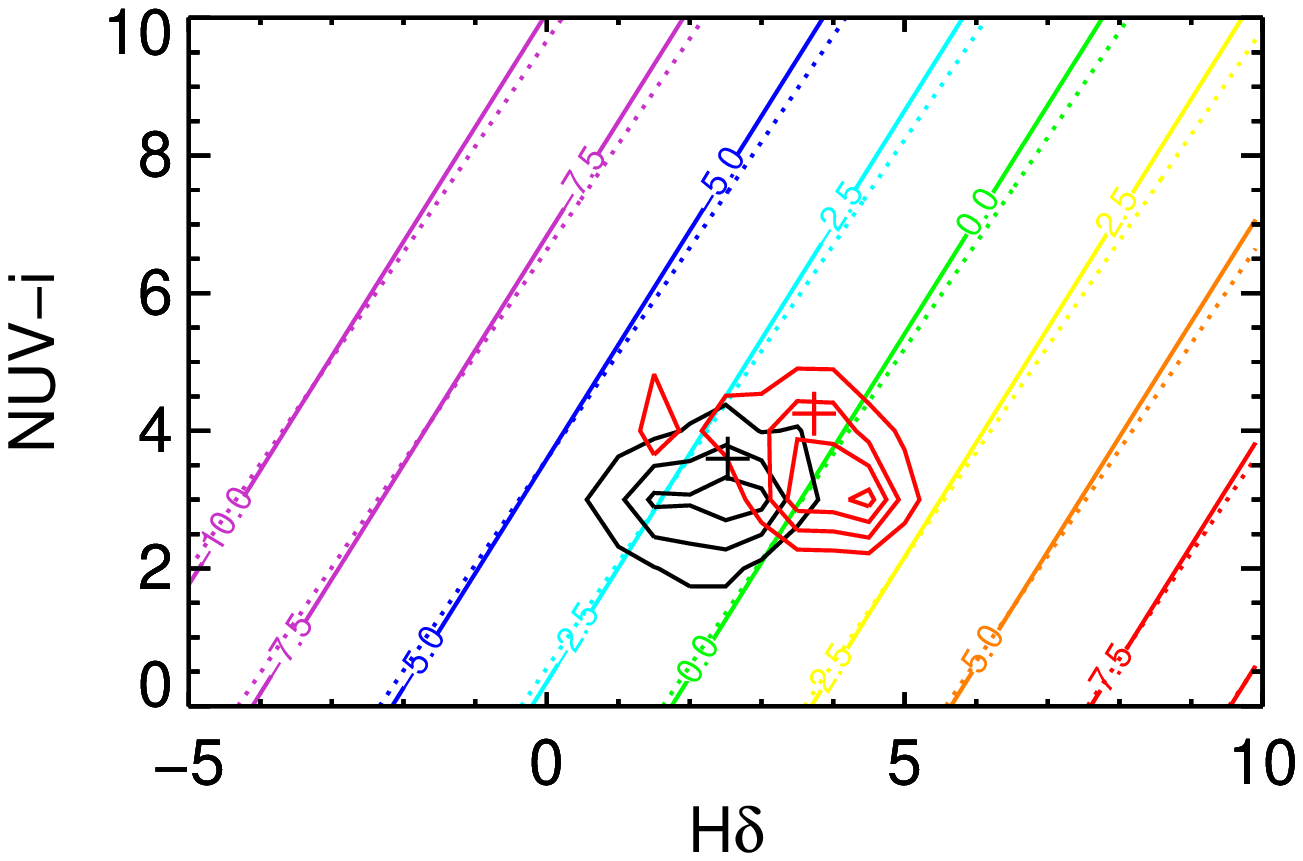}{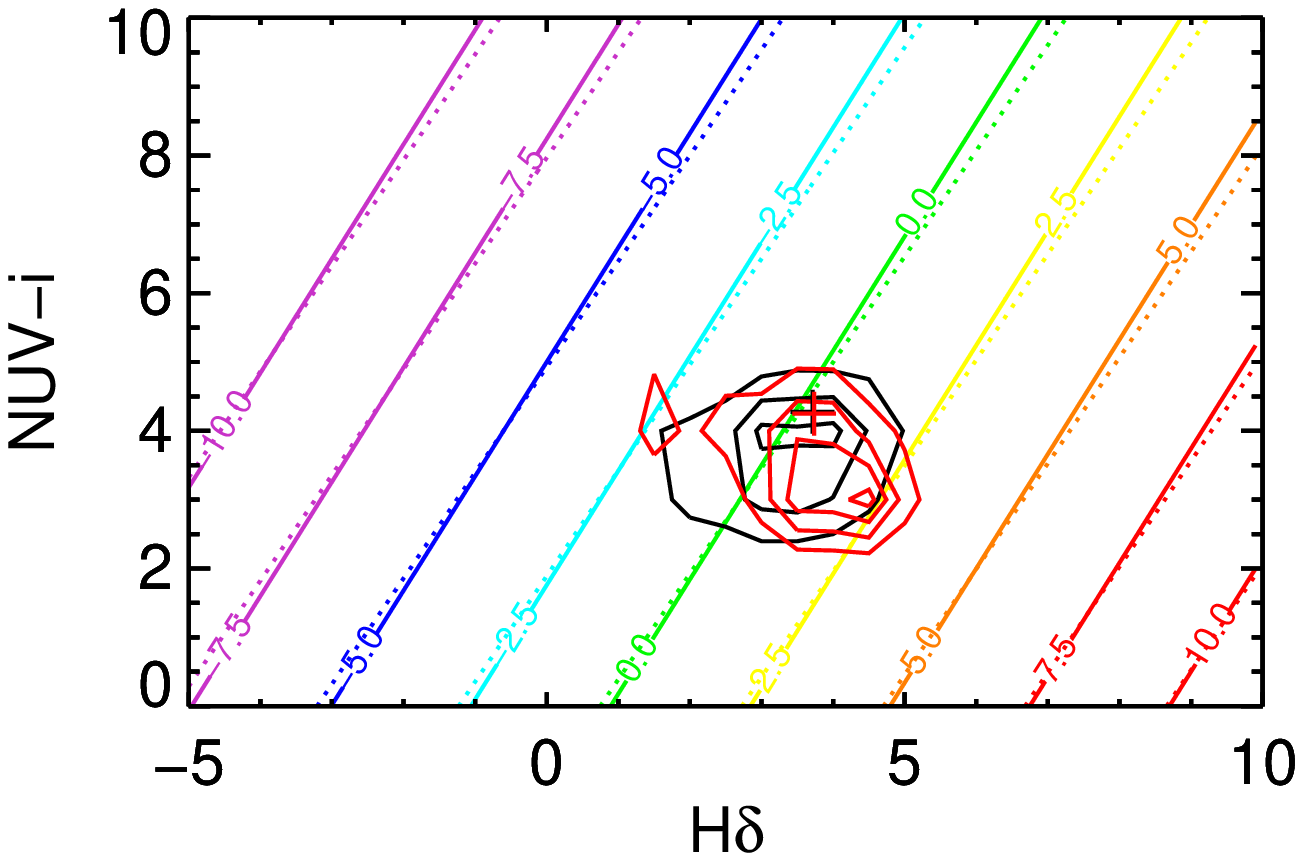}
\caption{a. Same as Figure \ref{fig_correct_nuvu_1.45} for NUV-i vs. \hda.
\label{fig_correct_nuvi_1.45}}
\end{figure}

\begin{figure}
\plotone{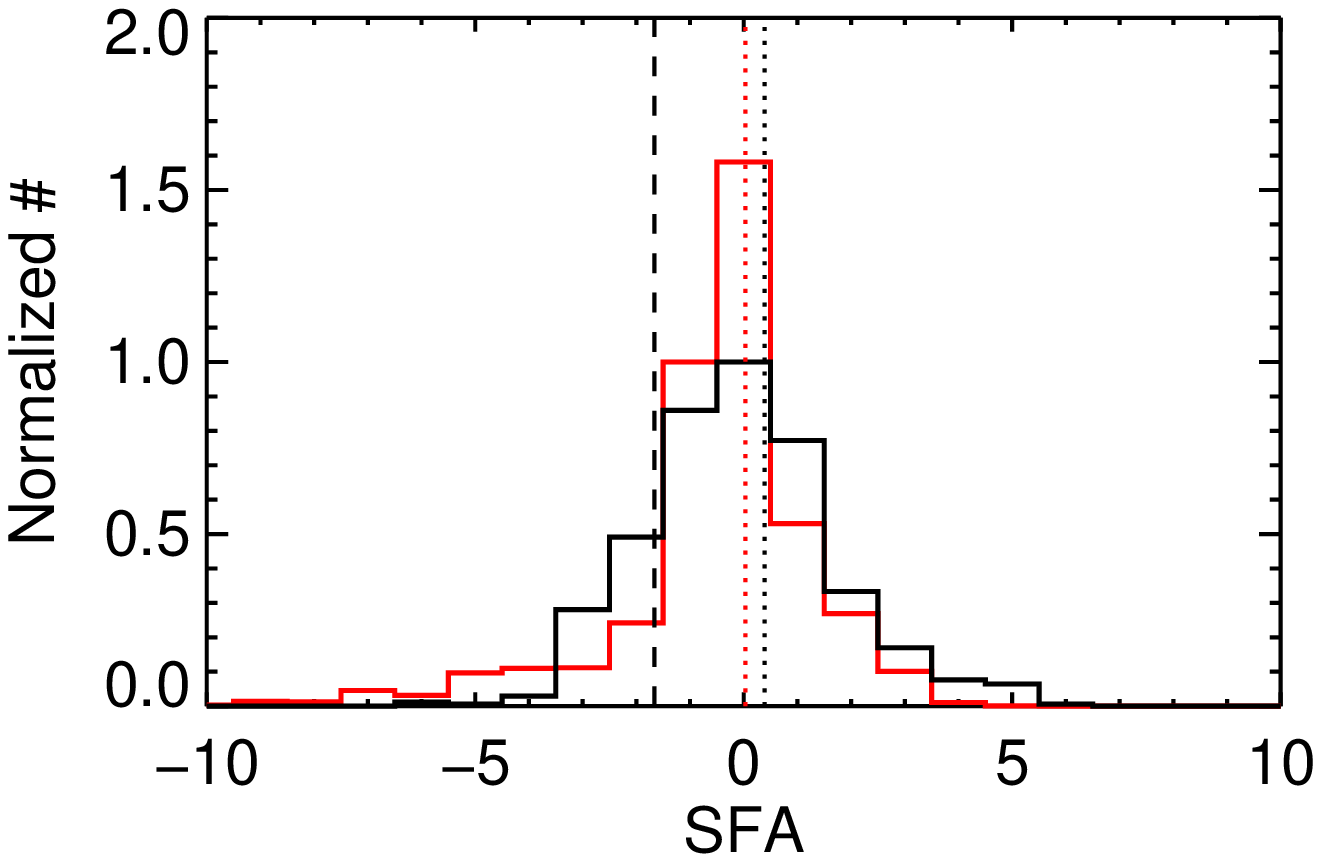}
\caption{Histogram of derived SFA (with corrected observables, black) vs. model SFA. Dotted lines show mean values (black: observable SFA, red: model SFA). Dashed line shows mean derived SFA for uncorrected observables. All in \dn=1.45 bin.
\label{fig_hist_sfa_1.45}}
\end{figure}

\begin{figure}
\plottwo{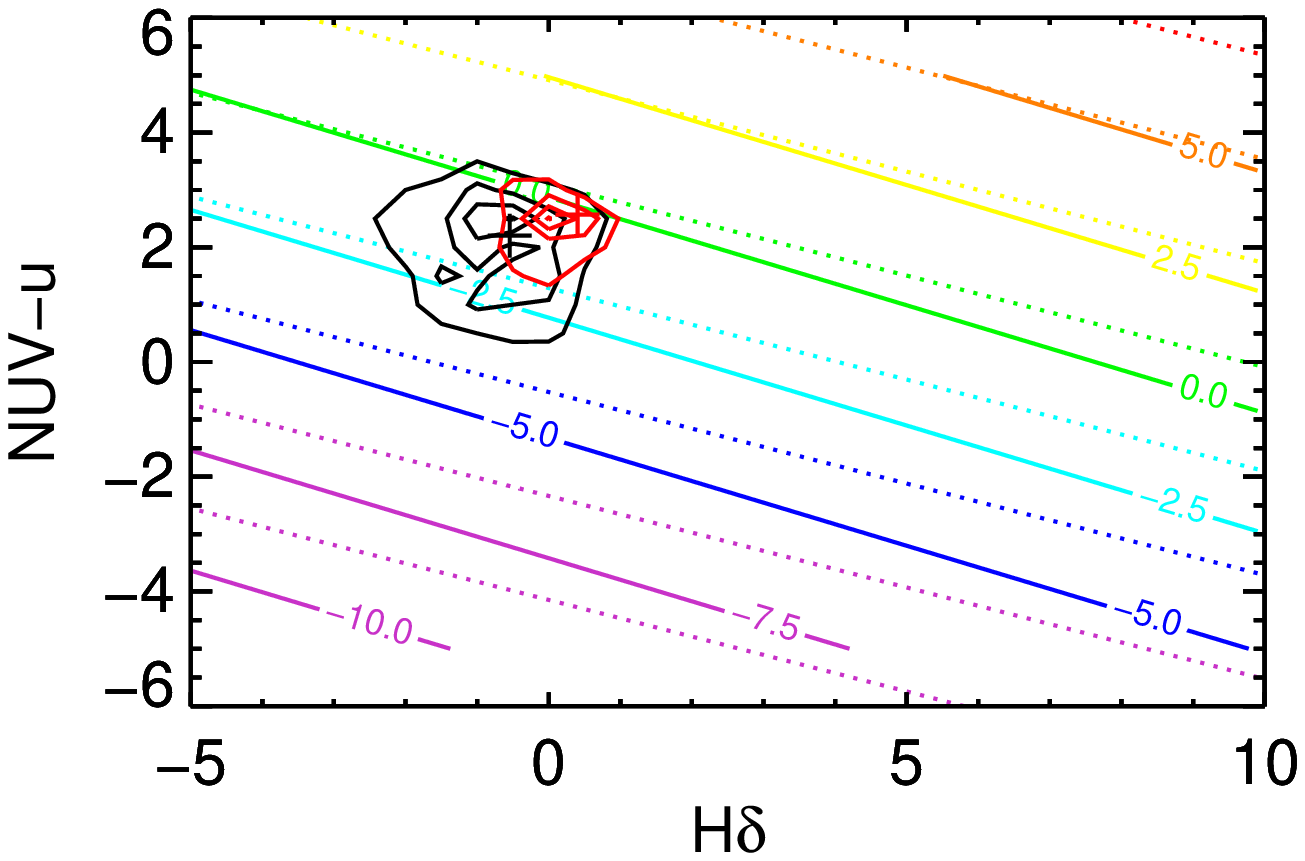}{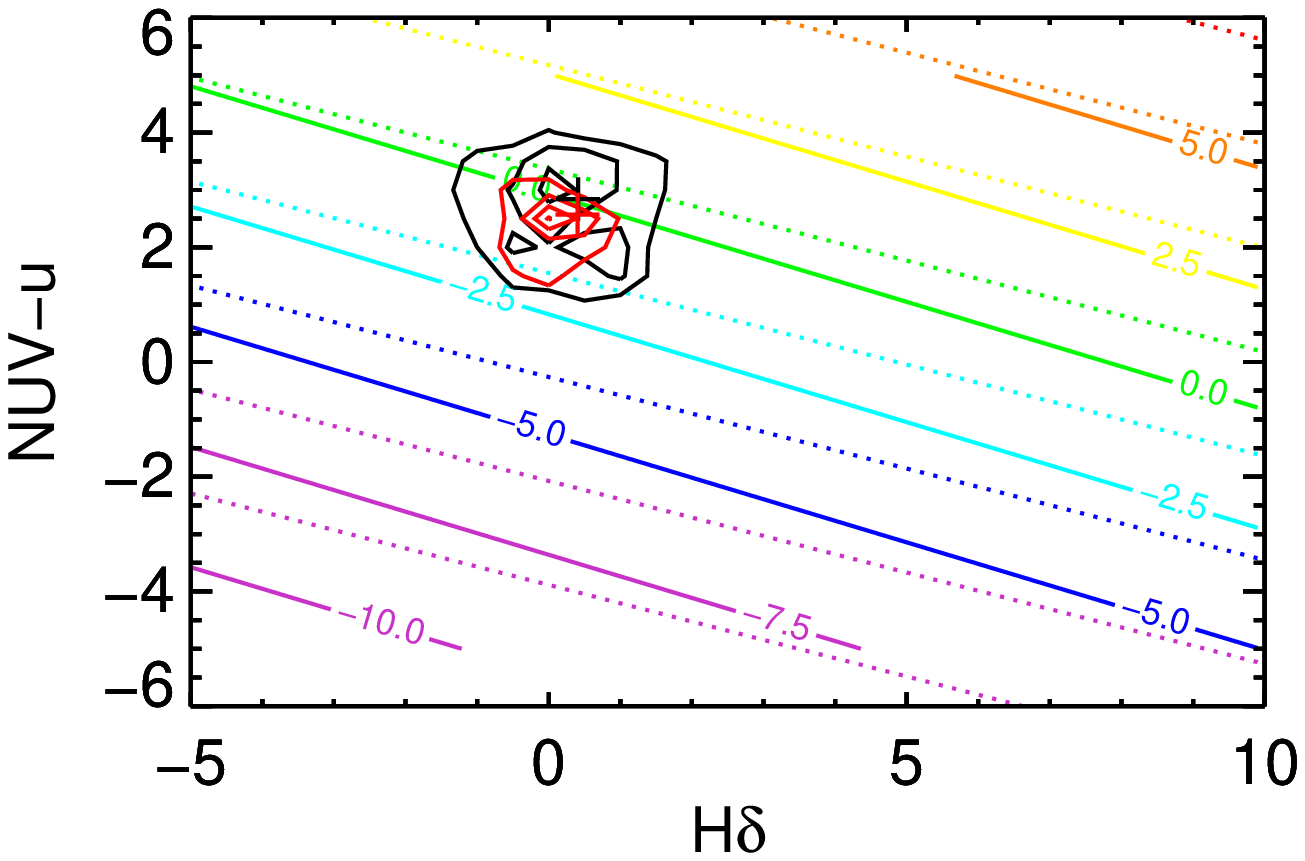}
\caption{a. Distribution of observables (\hda and NUV-u) in black contours, and model values in red contours, with means shown with crosses. For \dn=1.75 bin. Contours show SFA vs. \hda and NUV-u calculated using the mean values of the other observables in this \dn4 bin. Solid lines shows SFA calculated using FUV-NUV (or $\beta$), while dotted lines show SFA calculated not using FUV-NUV. b. Same as a. with corrected observables and SFA contours calculated using corrected mean observables.
\label{fig_correct_nuvu_1.75}}
\end{figure}

\begin{figure}
\plottwo{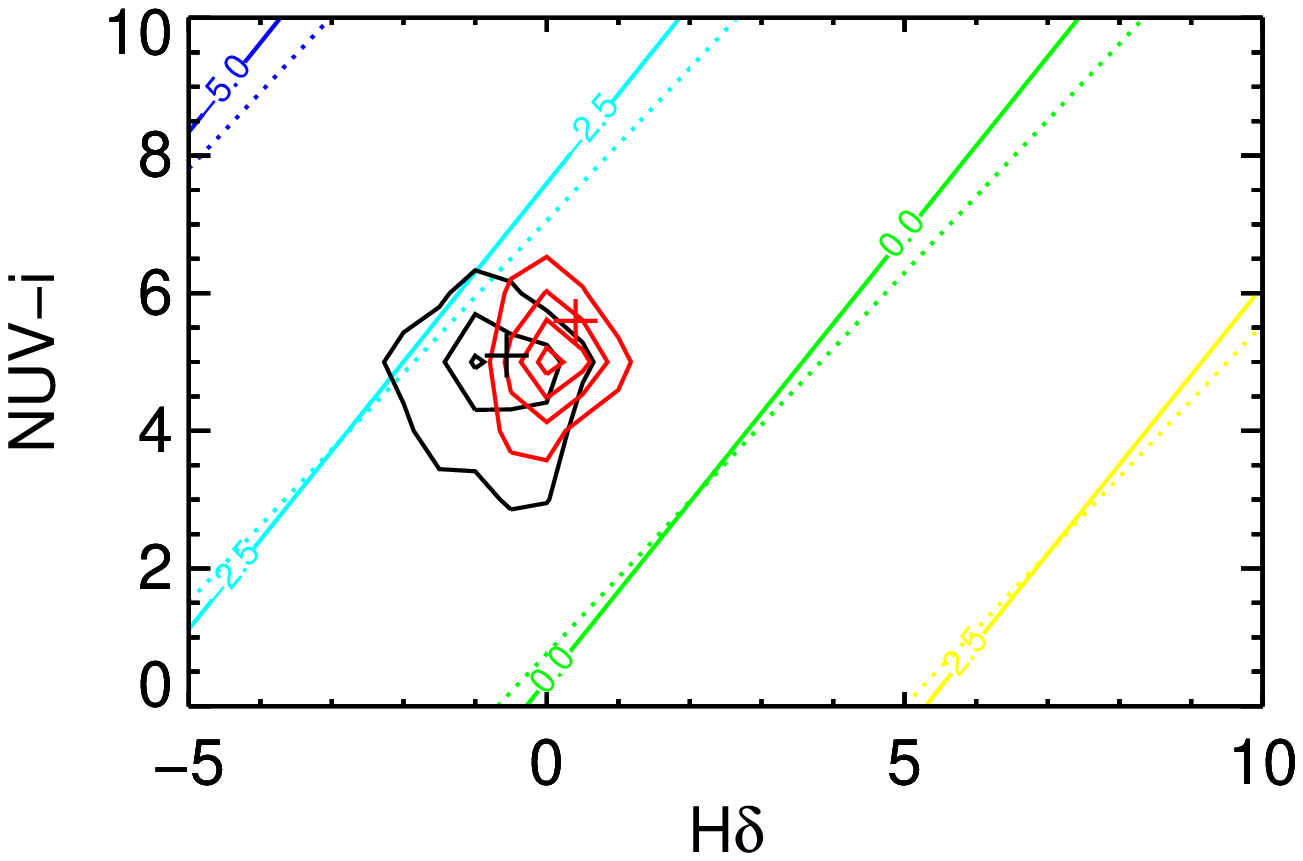}{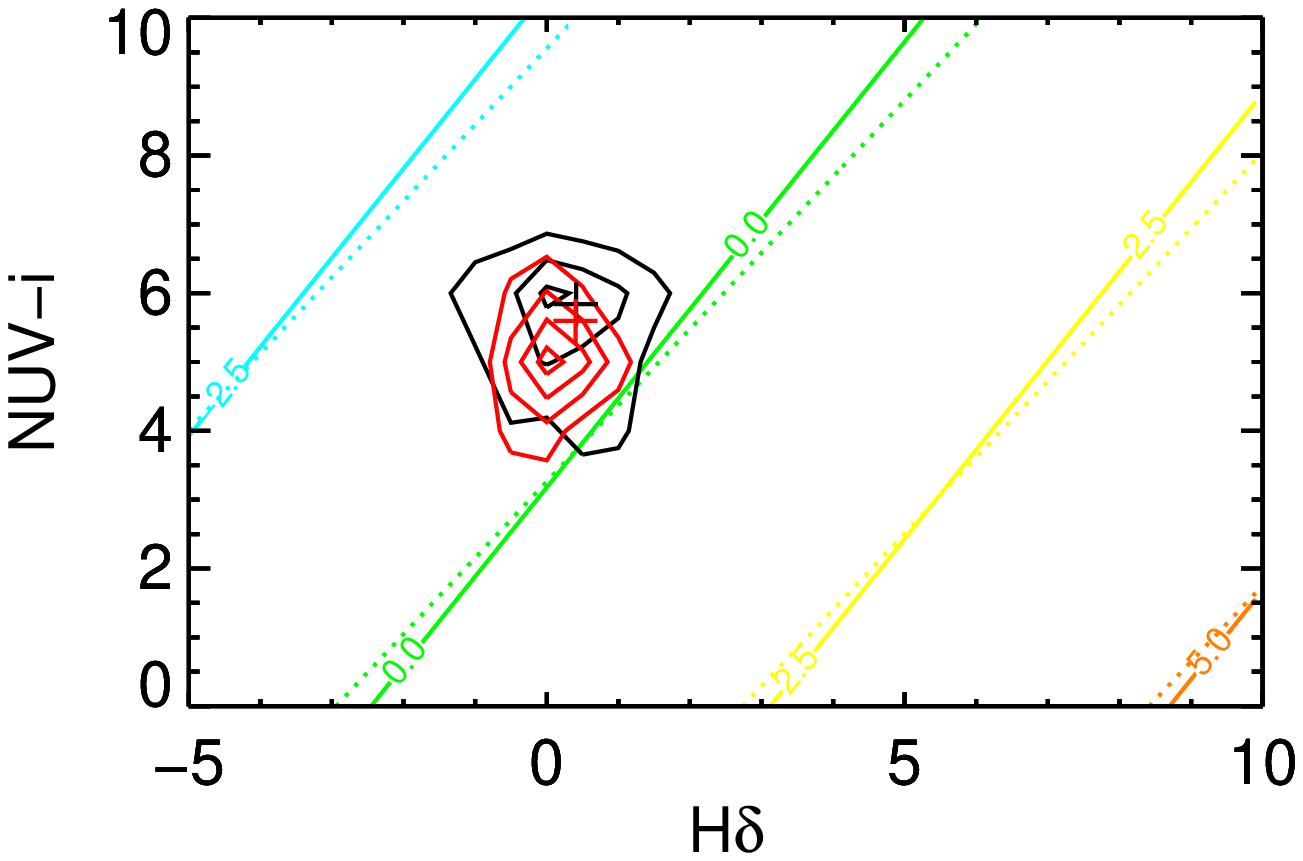}
\caption{a. Same as Figure \ref{fig_correct_nuvu_1.75} for NUV-i vs. \hda.
\label{fig_correct_nuvi_1.75}}
\end{figure}

\begin{figure}
\plotone{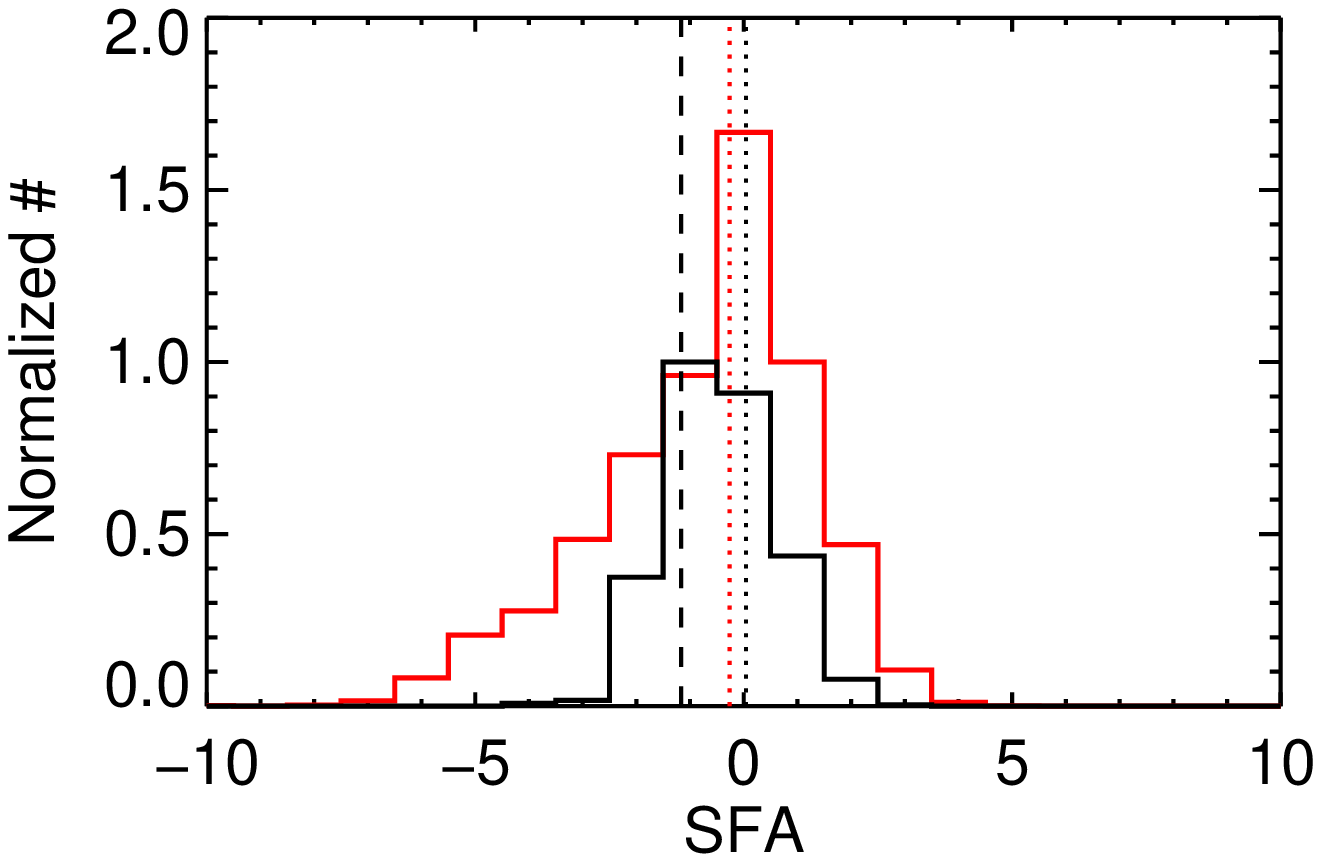}
\caption{Histogram of derived SFA (with corrected observables, black) vs. model SFA. Dotted lines show mean values (black: observable SFA, red: model SFA). Dashed line shows mean derived SFA for uncorrected observables. All in \dn=1.75 bin.
\label{fig_hist_sfa_1.75}}
\end{figure}

We give a few samples of the distributions in the next set of Figures. In Figure \ref{fig_correct_nuvu_1.25}, we show the distributions of observed NUV-u and \hda for the \dn=1.25 bin. In Figure \ref{fig_correct_nuvu_1.25}a, we show the uncorrected observed values vs. the model distribution. The plot also shows SFA contours using mean values for the other (observed) parameters, to show how variations in NUV-u and \hda affect SFA. In Figure \ref{fig_correct_nuvu_1.25}b we show the corrected observed values, model distribution, and SFA contours using the corrected mean observed values. See caption for further details. In Figure \ref{fig_correct_nuvi_1.25} we show the same information for NUV-i. We show the distributions of SFA in the \dn4=1.25 bin in Figure \ref{fig_hist_sfa_1.25}. We repeat these figures for \dn=1.45 (Figures \ref{fig_correct_nuvu_1.45}, \ref{fig_correct_nuvi_1.45}, \ref{fig_hist_sfa_1.45}); and for \dn=1.75 (Figures \ref{fig_correct_nuvu_1.75}, \ref{fig_correct_nuvi_1.75}, \ref{fig_hist_sfa_1.75}). The figures showing the SFA distributions illustrate that the observable adjustments bring the derived SFA into agreement with the model SFAs in their mean values. Without the corrections the two SFA distributions would be significantly discrepent. In general the spread in the derived SFA is similar to or higher than that in the model SFAs (Figures \ref{fig_hist_sfa_1.25},\ref{fig_hist_sfa_1.45}, \ref{fig_hist_sfa_1.75}).

Finally in Figure \ref{fig_sfa_flux_dsfa} we show a version of Figure \ref{fig_sfa_flux} with arrows added indicating how the observable corrections and associated SFA changes impact the flux diagram. There is a modest impact, typically moving the quench/burst point about 0.1 dex down in the quench direction in log SSFR.

We note as further evidence of the validity of this approach that the quenching rate derived in Table \ref{tab_method4} and Figure \ref{fig_rhobr} is consistent with the results of \cite{martin07}, which was obtained using an independent method. Both of these results are quanititatively consistent with the observed evolution in the galaxy main sequence and red sequences. Without reconciling the model and observation distributions, there would be a very significant discrepency between the derived quenching mass flux and the main and red sequence evolution.

\section{Comparison to Semi-Analytic Model Trends} \label{sec_model_comp}

\begin{figure}
\plotone{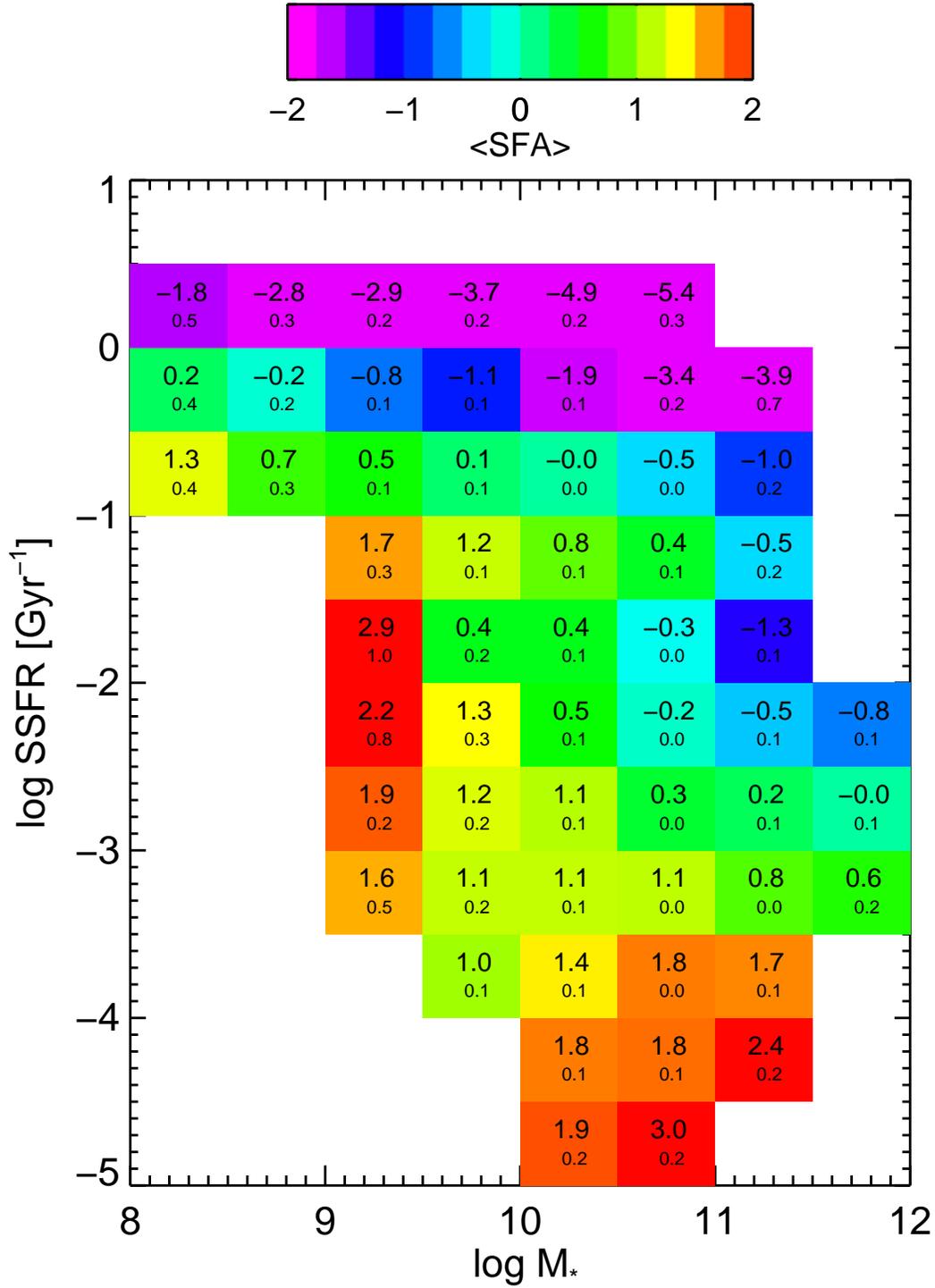}
\caption{Mean Star Formation Acceleration (SFA) indicated by color on the  log sSFR vs. log $M_*$ diagram for SDSS galaxies. Large number in box is mean SFA, small number is the standard error of the mean.
\label{fig_sfa_ssfr_mstar}}
\end{figure}

\begin{figure}
\plottwo{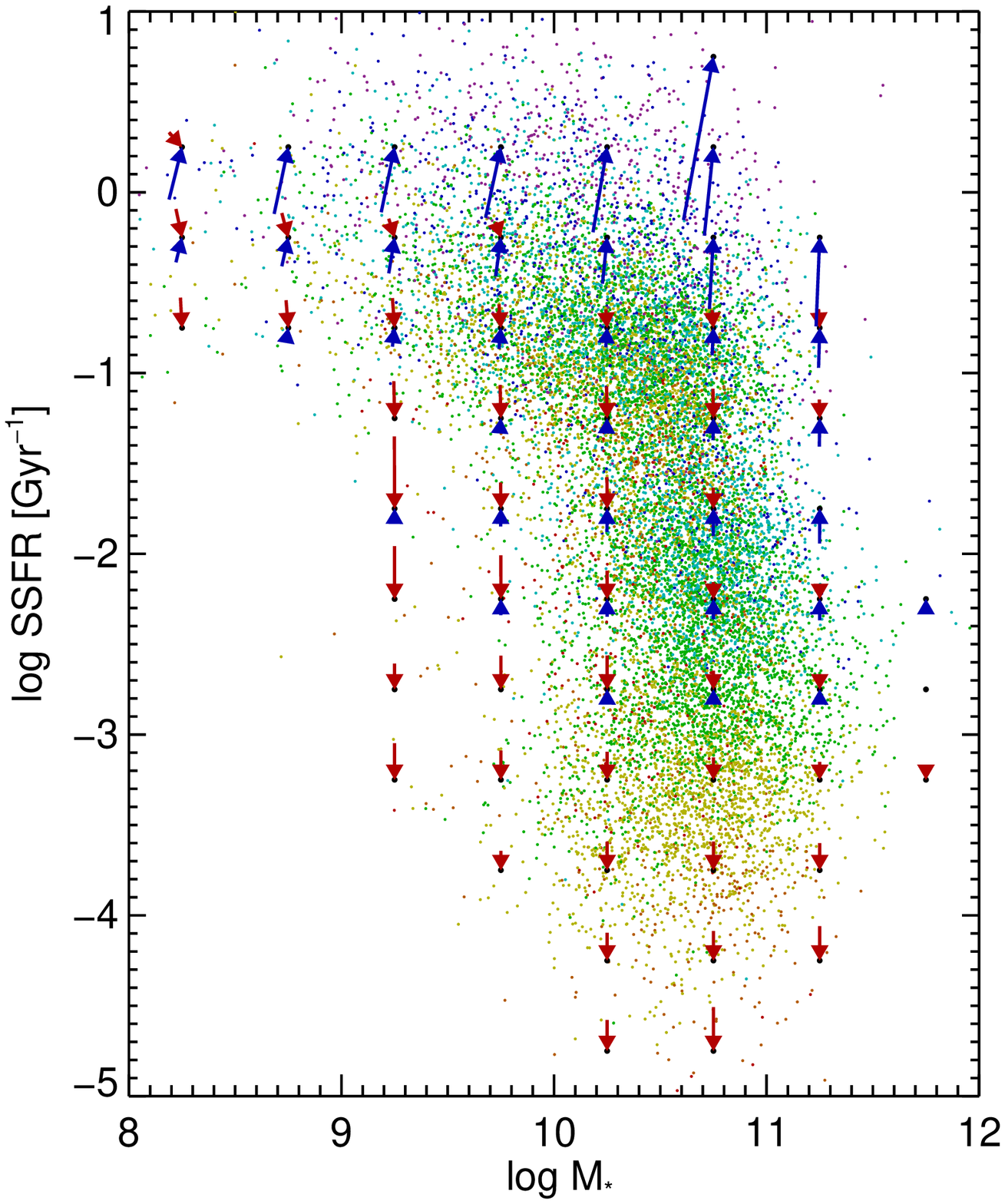}{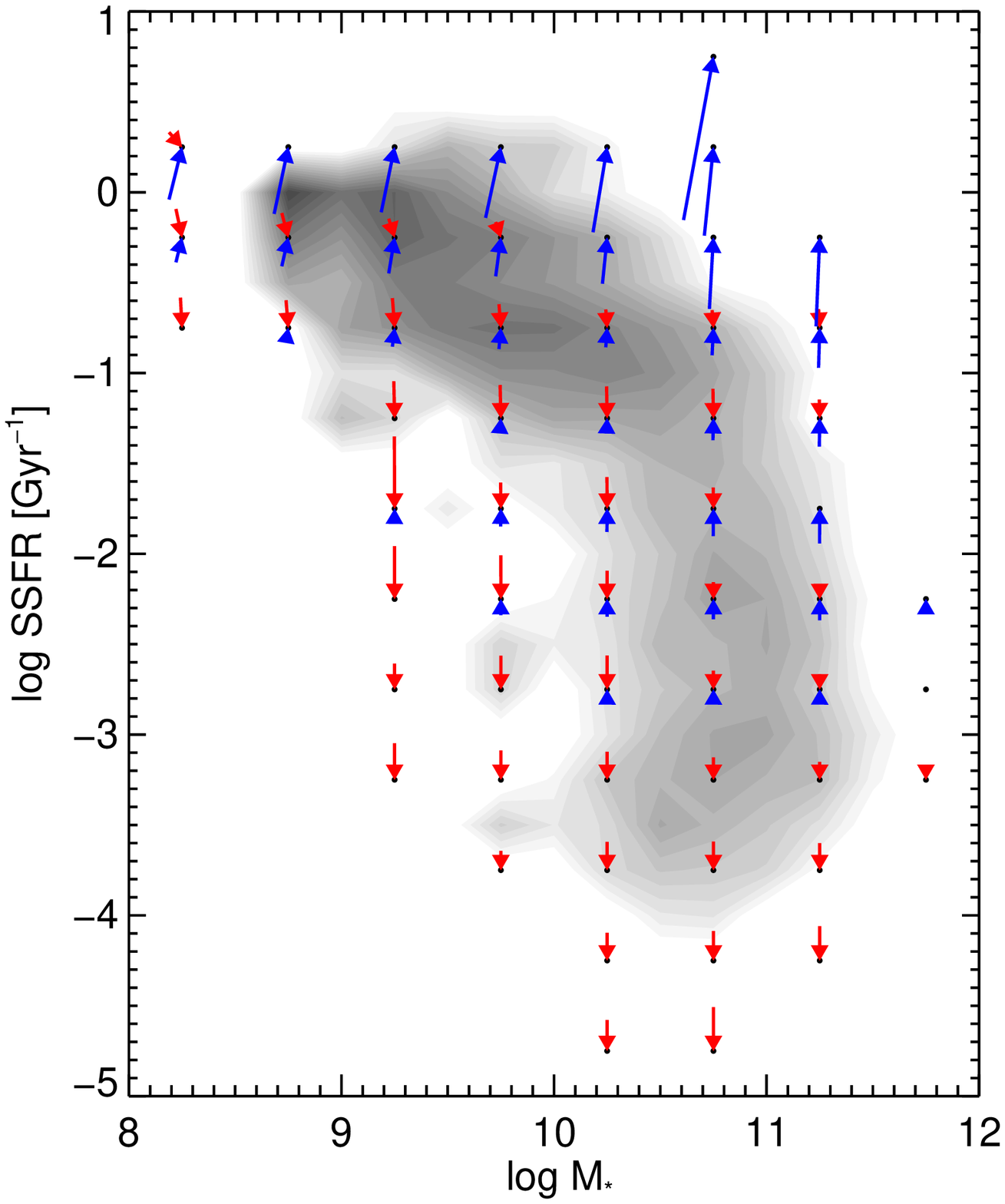}
\caption{a. Star Formation Acceleration (SFA) plotted as a flux vector on the sSFR vs. $M_*$ diagram for SDSS galaxies. There is a large spread in each bin, with galaxies that are bursting and those that are quenching in each bin. The purpose of this diagram is to try to represent this diversity using red (quenching) and blue (bursting) arrows whose length is roughly proportional to the typical quench and burst rate. These arrows then show the evolution of the average galaxy on the CMD, including effects of star formation (mass growth) and sSFR evolution (tracked by SFA). The length of the arrows is 1.5$\sigma_{SFA}$, 1.5 times the standard deviation of the SFA in that bin. Red and blue arrows give the relative amplitude of quenching and bursting respectively, with the length of red (blue) arrow equal to $1.5\sigma_{SFA}+\overline{SFA}$ ($1.5\sigma_{SFA}-\overline{SFA}$), multiplied by a factor that converts the SFA into $\Delta sSFR$ assuming a 100 Myr time interval. The head of each arrow is the current mass and sSFR, while the tail gives the typical point on the CMD where galaxies making up the current mass-sSFR were located 100 Myrs in the past. The sum of the two vectors is proportional to the average SFA in each bin displayed in Figure \ref{fig_sfa_ssfr_mstar}. 
The following trends are apparent. Blue galaxies have bursting SFAs, while red galaxies tend toward quenching SFA. Dots give individual galaxies colored by SFA (red: SFA=5, purple: SFA=-5). b. Same as a. with volume-corrected density plotted in greyscale contours. Volume correction as in M07. Levels are logarithmic with equal spacing between $5\times 10^{-5} < \phi [Mpc^{-3}] < 10^{-4}$, where volume density is per unit 0.5 dex bin in log $M_*$ and log SSFR.
\label{fig_sfa_flux}}
\end{figure}

\begin{figure}
\plotone{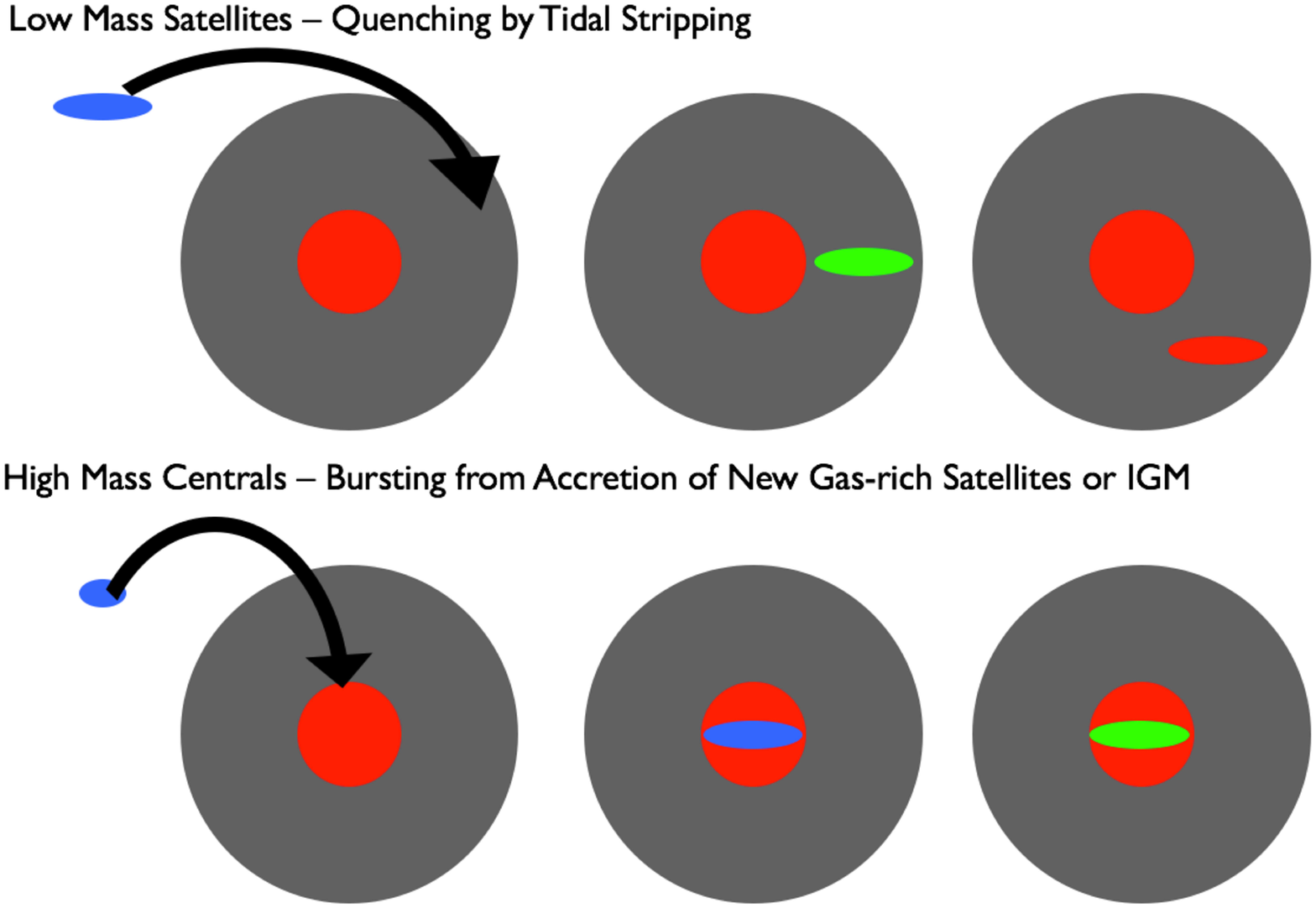}
\caption{Cartoon model for results displayed in Figures \ref{fig_sfa_nuvi}-\ref{fig_sfa_ssfr}. Low mass galaxies are stripped as they enter higher-mass halos and are tidally or ram pressure stripped of gas and quench. High mass galaxies are preferentially central galaxies and occasional suffer bursts from accretion events (merging satellites or infalling circum-galactic gas).
\label{fig_sfa_model}}
\end{figure}

\begin{figure}
\plotone{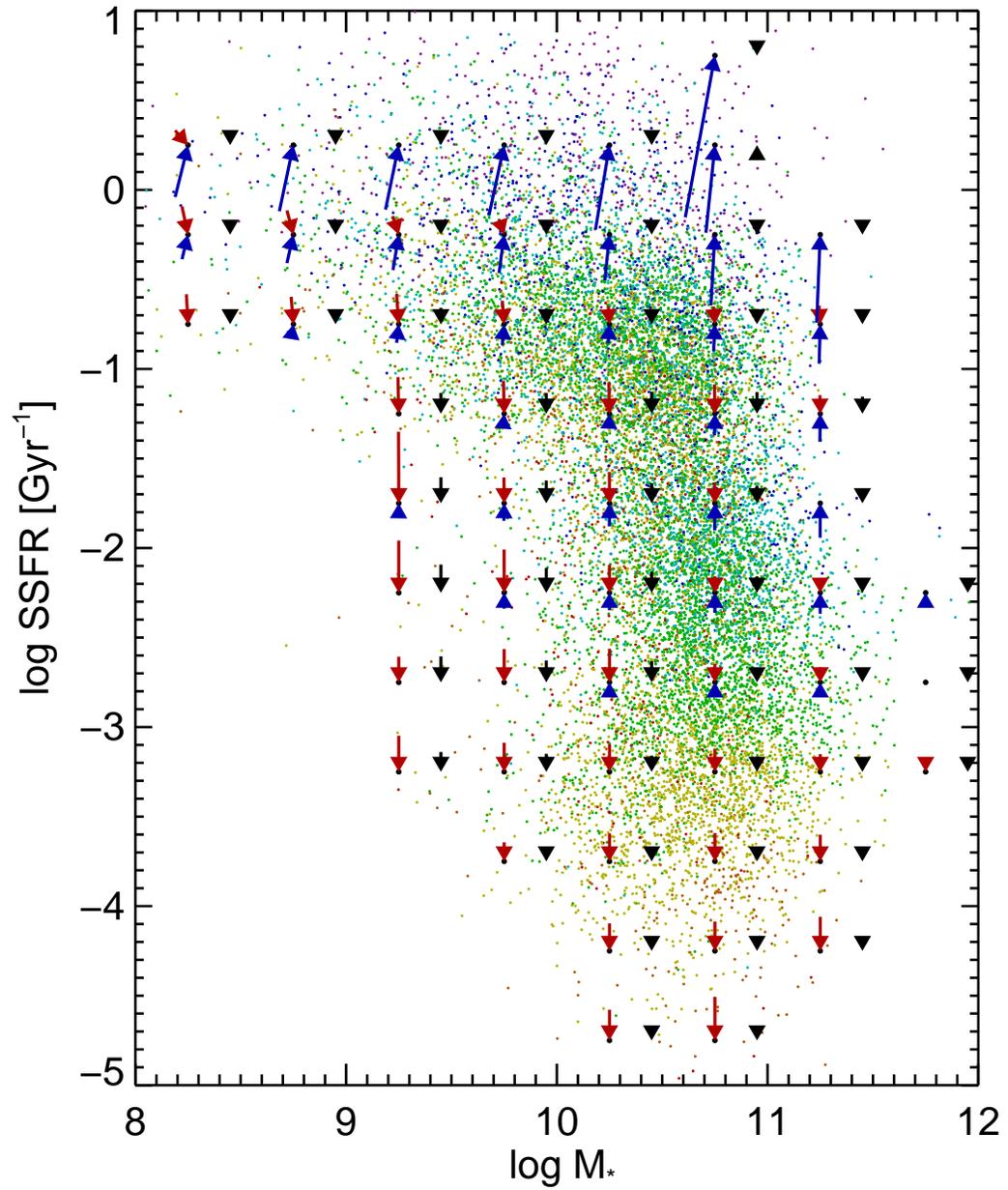}
\caption{Star Formation Acceleration (SFA) plotted as a flux vector on the sSFR vs. $M_*$ diagram for SDSS galaxies. Same as Figure \ref{fig_sfa_flux}. Added black arrows show change in quench/burst mean point produced by SFA change from observable distribution correction.
\label{fig_sfa_flux_dsfa}}
\end{figure}

\begin{figure}
\plottwo{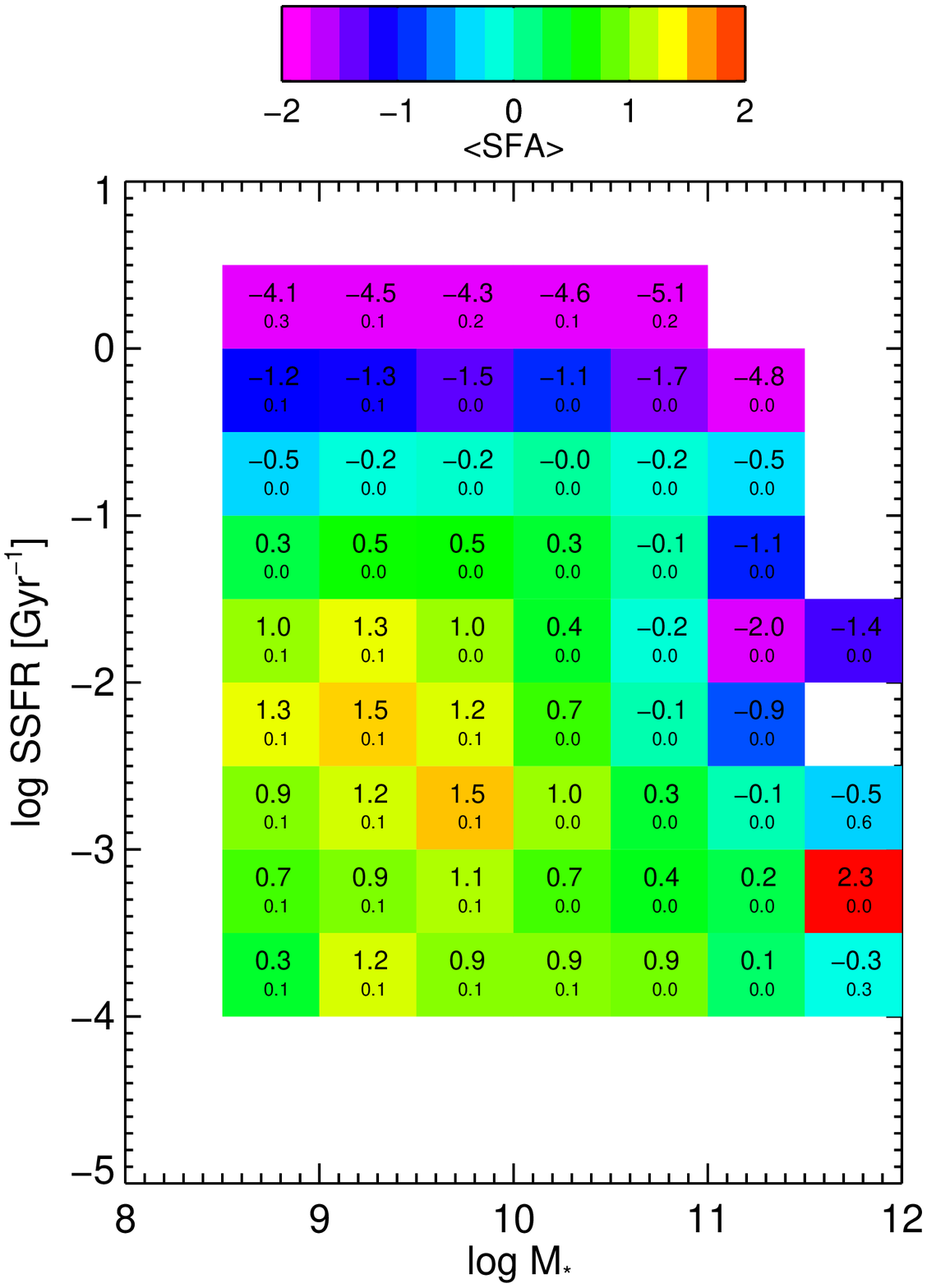}{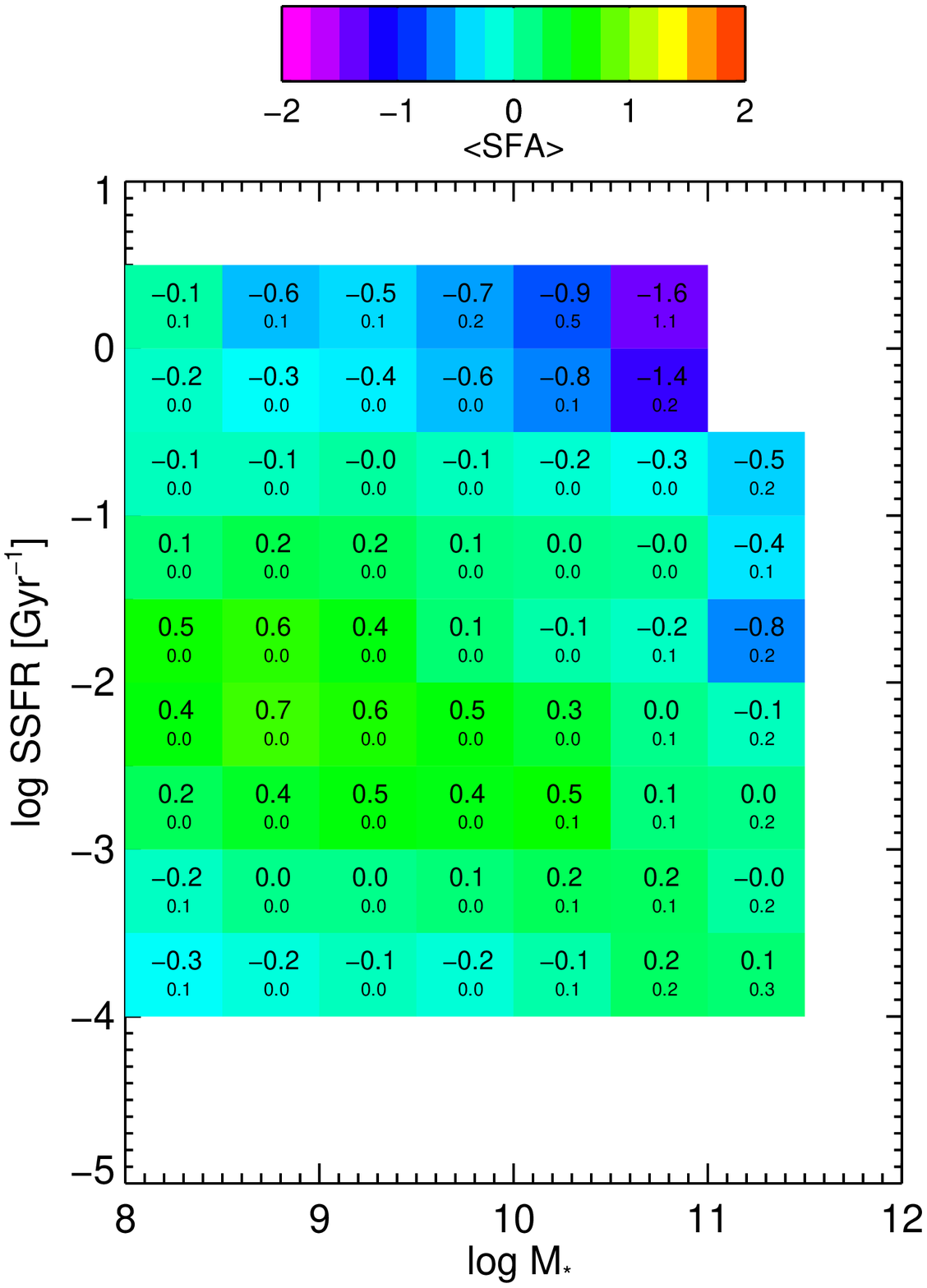}
\caption{a. Mean Star Formation Acceleration (SFA) indicated by color on the  log sSFR vs. log $M_*$ diagram for model galaxies.  Compare to Figure \ref{fig_sfa_ssfr_mstar}. b. Difference between model SFA and regression fit SFA ($<SFA>-<SFA_{fit}>$) averaged in each log sSFR-log $M_*$ bin. Maximum differences are typically 0.0-0.5 over most of the diagram. Correcting for these biases would slightly increase the mass trends discussed in the text.
\label{fig_sfa_ssfr_mstar_model}}
\end{figure}

\begin{figure}
\plotone{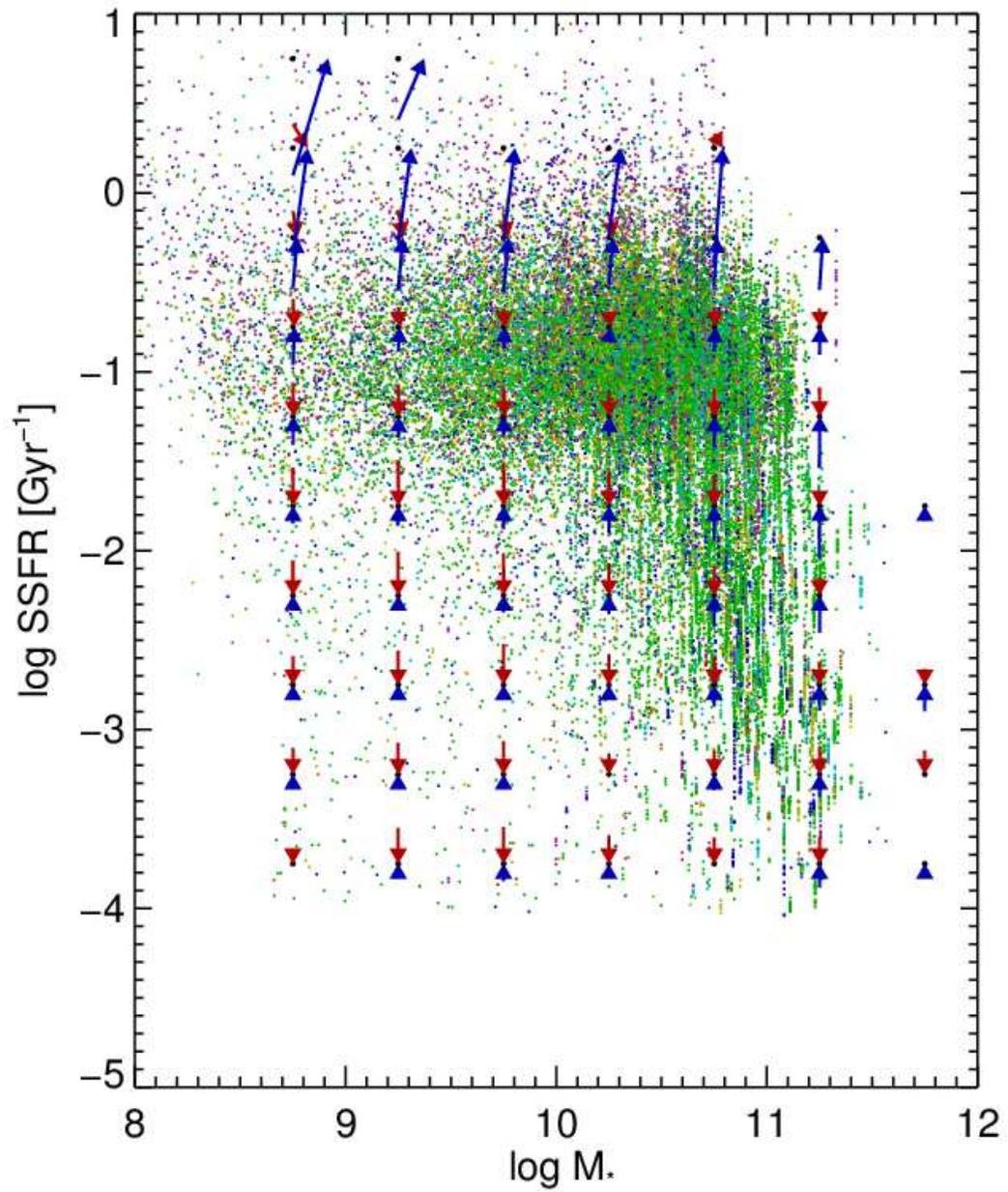}
\caption{Star Formation Acceleration (SFA) plotted as a flux vector on the sSFR vs. $M_*$ diagram for model galaxies. Compare to Figure \ref{fig_sfa_flux}.
\label{fig_sfa_flux_model}}
\end{figure}

We have repeated the analysis of \S\ref{sec_app1} for the model galaxies used to generate the fitting coefficients. We choose galaxies in the redshift range $0<z<0.3$ using the magnitude-limited subset discussed above. The mean SFA over this diagram is given in Figure \ref{fig_sfa_ssfr_mstar_model}a (the equivalent of Figure \ref{fig_sfa_ssfr_mstar}.  We also show the impact of fitting error on this diagram in Figure \ref{fig_sfa_ssfr_mstar_model}b. This shows that biases introduced by fitting SFA are typically 0.0-0.5. Correcting for these small biases on this diagram would slightly amplify the observed trends. In Figure \ref{fig_sfa_flux_model} we show the flux diagram that is the equivalent of Figure \ref{fig_sfa_flux}. The trends with SSFR (SFA decreasing) and with mass (SFA decreasing with increasing mass in the green valley) are similar, but the amplitudes are smaller. When we perform the equivalent of the calculation of Table \ref{tab_method4}, we find a mass flux (over the same mass range) of $\dot{\rho}_{BR}=(3.3 \times 10^{-4} M_\odot yr^{-1} Mpc^{-3}$, a factor of $\sim$100 lower than our observed mass flux and that inferred from the evolution of the blue and red galaxy luminosity functions.






\clearpage
\begin{deluxetable}{lccccccccc}
\tabletypesize{\scriptsize}
\tablecaption{Regression Coefficients Sample
D$_n$(4000)=1.30  0.0$<$z$<$0.3
\label{tab_coef}}
\tablewidth{0pt}
\tablehead{
\colhead{Parameter}&
\colhead{$\beta$}&
\colhead{NUV-u}&
\colhead{u-g}&
\colhead{g-r}&
\colhead{NUV-i}&
\colhead{D$_n$(4000)}&
\colhead{H$\delta_a$}&
\colhead{M$_i$}&
\colhead{Const}
}
\startdata
     A$_{FUV}$&  -0.055&  -0.606&   0.496&   0.530&   0.806&  -0.456&   0.029&  -0.034&  -0.482\\
   (NUV-r)$_0$&  -0.113&  -0.281&   0.274&   0.334&   0.559&  -0.340&   0.018&  -0.037&   0.037\\
           SFA&   0.430&   1.977&   0.168&  -2.063&  -0.758&   1.673&   1.102&  -0.015&  -5.552\\
           SFJ&   0.114&   0.314&  -0.152&  -0.406&  -0.085&   0.597&   0.132&   0.013&  -1.673\\
      log(SFR)&  -0.259&  -0.846&   0.388&   0.568&   0.502&  -0.454&  -0.025&  -0.442&  18.630\\
     log(sSFR)&  -0.244&  -0.616&   0.255&   0.332&   0.279&  -0.436&   0.016&  -0.027&   0.903\\
    log(M$_*$)&  -0.015&  -0.230&   0.133&   0.236&   0.223&  -0.018&  -0.041&  -0.415&  17.727\\
      log(Age)&  -0.007&  -0.075&   0.013&   0.084&   0.027&   0.032&  -0.030&  -0.010&   1.247\\
log(M$_{gas}$)&  -0.042&  -0.061&   0.045&   0.041&   0.065&  -0.113&  -0.055&  -0.255&  11.201\\
     Z$_{gas}$&  -0.039&  -0.143&   0.114&   0.115&   0.137&  -0.040&  -0.035&  -0.336&  12.459\\
         Z$_*$&  -0.053&  -0.428&   0.250&   0.369&   0.331&   0.021&  -0.041&  -0.504&  19.218\\
           Ext&   0.869&   3.710&   0.407&  -3.735&  -1.508&   2.669&   1.882&  -0.007&  -9.502\\
\enddata
\end{deluxetable}

\begin{deluxetable}{rrrrrrrrr}
\tablecaption{Influence Function
\label{tab_influence}}
\tablewidth{0pt}
\tablehead{
\colhead{Parameter}&
\colhead{$\beta$}&
\colhead{NUV-u}&
\colhead{u-g}&
\colhead{g-r}&
\colhead{NUV-i}&
\colhead{D$_n$(4000)}&
\colhead{H$\delta_a$}&
\colhead{M$_i$}
}
\startdata
     log(M$_*$)&      0.13&      0.14&      0.12&      0.10&      0.13&      0.00&      0.08&      0.88\\
      A$_{FUV}$&      0.12&      0.29&      0.32&      0.31&      0.38&      0.00&      0.01&      0.02\\
$\Delta$(NUV-i)&      0.13&      0.31&      0.31&      0.31&      0.38&      0.00&      0.01&      0.04\\
       log(SFR)&      0.34&      0.23&      0.16&      0.16&      0.18&      0.00&      0.12&      0.69\\
      log(sSFR)&      0.49&      0.28&      0.16&      0.17&      0.20&      0.01&      0.14&      0.22\\
            SFA&      0.17&      0.16&      0.03&      0.14&      0.10&      0.00&      0.36&      0.05\\
            SFJ&      0.00&      0.00&      0.00&      0.00&      0.00&      0.01&      0.16&      0.07\\
       log(Age)&      0.04&      0.02&      0.01&      0.03&      0.01&      0.01&      0.16&      0.10\\
 log(M$_{gas}$)&      0.09&      0.05&      0.00&      0.00&      0.01&      0.00&      0.05&      0.55\\
      Z$_{gas}$&      0.10&      0.08&      0.05&      0.03&      0.07&      0.00&      0.09&      0.73\\
          Z$_*$&      0.13&      0.14&      0.13&      0.11&      0.13&      0.00&      0.10&      0.86\\
\enddata
\end{deluxetable}

\begin{deluxetable}{lrrrrrrrrrrr}
\tablecaption{Degeneracy Function
\label{tab_degeneracy}}
\tablewidth{0pt}
\tablehead{
\colhead{Parameter}&
\colhead{A$_{FUV}$}&
\colhead{$\Delta$(NUV-i)$_0$}&
\colhead{SFA}&
\colhead{SFJ}&
\colhead{log(SFR)}&
\colhead{log(sSFR)}&
\colhead{log(M$_*$)}&
\colhead{log(Age)}&
\colhead{log(M$_{gas}$)}&
\colhead{Z$_{gas}$}&
\colhead{Z$_*$}
}
\startdata
      A$_{FUV}$&  1.00&  0.98& -0.68& -0.54&  0.87&  0.86&  0.61&  0.25&  0.17&  0.40&  0.66\\
$\Delta$(NUV-i)& -1.00&  1.00& -0.64& -0.53&  0.84&  0.82&  0.59&  0.20&  0.19&  0.41&  0.63\\
            SFA& -1.00& -1.00&  1.00&  0.62& -0.74& -0.76& -0.46& -0.43& -0.16& -0.25& -0.47\\
            SFJ& -1.00& -1.00& -1.00&  1.00& -0.56& -0.60& -0.31& -0.20& -0.25& -0.30& -0.33\\
       log(SFR)& -1.00& -1.00& -1.00& -1.00&  1.00&  0.92&  0.79&  0.29&  0.37&  0.58&  0.81\\
      log(sSFR)& -1.00& -1.00& -1.00& -1.00& -1.00&  1.00&  0.50&  0.16&  0.20&  0.35&  0.55\\
     log(M$_*$)& -1.00& -1.00& -1.00& -1.00& -1.00& -1.00&  1.00&  0.42&  0.56&  0.79&  0.98\\
       log(Age)& -1.00& -1.00& -1.00& -1.00& -1.00& -1.00& -1.00&  1.00&  0.01&  0.19&  0.47\\
 log(M$_{gas}$)& -1.00& -1.00& -1.00& -1.00& -1.00& -1.00& -1.00& -1.00&  1.00&  0.88&  0.49\\
      Z$_{gas}$& -1.00& -1.00& -1.00& -1.00& -1.00& -1.00& -1.00& -1.00& -1.00&  1.00&  0.77\\
          Z$_*$& -1.00& -1.00& -1.00& -1.00& -1.00& -1.00& -1.00& -1.00& -1.00& -1.00&  1.00\\
\enddata
\end{deluxetable}

\begin{deluxetable}{rrrrccrr}
\tablecaption{Mass Flux Table Method 4 \label{tab_method4}}
\tablewidth{0pt}
\tablehead{
\colhead{M$_r$} & 
\colhead{$\log{M_*}$} & 
\colhead{$\phi$}&
\colhead{$\#$}  & 
\colhead{$\overline{dy/dt}$[M07]} & 
\colhead{$\overline{dy/dt}$} & 
\colhead{$\dot{\rho}_{BR}$} & 
\colhead{$\sigma[\dot{\rho}_{BR}]$}
}
\startdata
  -23.75 &    11.63 &   1.84e-07 &     5 &     0.82 &     0.82 &    1.3e-04 &    1.8e-05\\
  -23.25 &    11.49 &   1.08e-06 &    22 &     0.70 &     0.69 &    4.6e-04 &    6.2e-05\\
  -22.75 &    11.32 &   6.53e-06 &   100 &     0.85 &     0.54 &    1.5e-03 &    1.9e-04\\
  -22.25 &    11.12 &   2.20e-05 &   217 &     0.87 &     0.53 &    3.1e-03 &    3.2e-04\\
  -21.75 &    10.93 &   4.02e-05 &   252 &     0.94 &     0.61 &    4.2e-03 &    3.2e-04\\
  -21.25 &    10.73 &   5.63e-05 &   211 &     0.87 &     0.69 &    4.2e-03 &    2.5e-04\\
  -20.75 &    10.50 &   7.42e-05 &   154 &     1.05 &     0.81 &    3.8e-03 &    2.8e-04\\
  -20.25 &    10.25 &   6.80e-05 &    77 &     1.37 &     0.93 &    2.3e-03 &    2.0e-04\\
  -19.75 &    10.00 &   6.94e-05 &    40 &     1.69 &     1.17 &    1.6e-03 &    2.5e-04\\
  -19.25 &     9.85 &   6.06e-05 &    21 &     1.50 &     1.32 &    1.1e-03 &    2.0e-04\\
  -18.75 &     9.52 &   7.22e-05 &     9 &     2.90 &     1.22 &    5.8e-04 &    1.4e-04\\
\tableline
Sum & & & & & &2.3e-02 & 7.4e-04\\
\enddata
\end{deluxetable}

\begin{deluxetable}{lcccccc}
\tablecaption{Color/Index Corrections \label{tab_colcor}}
\tablewidth{0pt}
\tablehead{
\colhead{D$_n$(4000)}&
\colhead{$\beta$}&
\colhead{NUV-u}&
\colhead{u-g}&
\colhead{g-r}&
\colhead{NUV-i}&
\colhead{H$\delta_a$}
}
\startdata
 1.10& 0.43& 0.25& 0.17& 0.20& 0.10& 0.62\\
 1.15& 0.48& 0.28& 0.12& 0.08& 0.06& 0.80\\
 1.20& 0.35& 0.25& 0.12& 0.07& 0.06& 0.90\\
 1.25& 0.32& 0.25& 0.13& 0.07& 0.06& 1.12\\
 1.30& 0.33& 0.26& 0.13& 0.06& 0.06& 1.39\\
 1.35& 0.47& 0.35& 0.13& 0.07& 0.07& 1.42\\
 1.40& 0.71& 0.43& 0.13& 0.05& 0.06& 1.19\\
 1.45& 1.08& 0.63& 0.11& 0.06& 0.06& 0.77\\
 1.50& 1.17& 0.71& 0.15& 0.05& 0.07& 0.67\\
 1.55& 1.32& 0.79& 0.11& 0.04& 0.06& 0.73\\
 1.60& 1.71& 0.90& 0.10& 0.02& 0.07& 0.74\\
 1.65& 1.37& 0.76& 0.09& 0.03& 0.06& 0.83\\
 1.70& 1.34& 0.62& 0.07& 0.00& 0.06& 0.97\\
 1.75& 1.56& 0.64& 0.02&-0.01& 0.06& 1.12\\
 1.80& 1.13& 0.41& 0.06&-0.01& 0.05& 1.13\\
 1.85& 0.96& 0.28&-0.03&-0.02& 0.07& 0.95\\
 1.90& 1.49& 0.53& 0.05& 0.06& 0.08& 1.34\\
 1.95& 0.69&-0.00& 0.06&-0.02& 0.04& 0.96\\
\tableline
\enddata
\end{deluxetable}


\clearpage



\clearpage

\clearpage

\end{document}


\end{document}
